\documentclass[conference,compsoc]{IEEEtran}
\ifCLASSOPTIONcompsoc
  \usepackage[nocompress]{cite}
\else
  \usepackage{cite}
\fi
\ifCLASSINFOpdf
  \usepackage[pdftex]{graphicx}
\else
\fi
\usepackage{amsmath}
\usepackage{url}
\usepackage{algorithm}
\usepackage{color}
\usepackage{comment}
\usepackage{listings}
\usepackage{booktabs}
\usepackage{multirow}

\usepackage{tikz}
\usepackage{amsmath}

\usepackage{filecontents}
\usepackage{comment}
\usepackage{booktabs}
\usepackage{algorithm}
\usepackage{algorithmic}
\usepackage{multirow}
\usepackage{listings}
\usepackage{xcolor}
\newcommand{\mj}[1] {{\footnotesize\color{red   }[MJ: #1]}}

\definecolor{mGreen}{rgb}{0,0.6,0}
\definecolor{mGray}{rgb}{0.5,0.5,0.5}
\definecolor{mPurple}{rgb}{0.58,0,0.82}

\lstdefinestyle{CStyle}{
    backgroundcolor=\color{white},   
    commentstyle=\color{mGreen},
    keywordstyle=\color{blue},
    numberstyle=\tiny\color{mGray},
    stringstyle=\color{mPurple},
    basicstyle=\footnotesize,
    frame=single,
    breakatwhitespace=false,         
    breaklines=true,                 
    captionpos=b,                    
    keepspaces=true,                 
    numbers=left,                    
    numbersep=5pt,                  
    showspaces=false,                
    showstringspaces=false,
    showtabs=false,                  
    tabsize=2,
    language=C
}

\begin{document}
%
% paper title
% Titles are generally capitalized except for words such as a, an, and, as,
% at, but, by, for, in, nor, of, on, or, the, to and up, which are usually
% not capitalized unless they are the first or last word of the title.
% Linebreaks \\ can be used within to get better formatting as desired.
% Do not put math or special symbols in the title.
\title{Causative Insights into Open Source Software Security using \\ Large Language Code Embeddings and Semantic Vulnerability Graph}

% Desc is part of repair

% author names and affiliations
% use a multiple column layout for up to three different
% affiliations

\author{\IEEEauthorblockN{Nafis Tanveer Islam}
\IEEEauthorblockA{Department of Computer Science\\
University of Texas at San Antonio\\
}
\and
\IEEEauthorblockN{Gonzalo De La Torre Parra}
\IEEEauthorblockA{Department of Cyber Security Systems\\
University of the Incarnate Word\\}
\and
\IEEEauthorblockN{Dylan Manuel}
\IEEEauthorblockA{Department of Computer Science\\
University of Texas at San Antonio\\
}
\and
\IEEEauthorblockN{Murtuza Jadliwala}
\IEEEauthorblockA{Department of Computer Science\\
University of Texas at San Antonio\\}
\and
\IEEEauthorblockN{Peyman Najafirad}
\IEEEauthorblockA{Department of Computer Science\\
University of Texas at San Antonio\\}
}

% conference papers do not typically use \thanks and this command
% is locked out in conference mode. If really needed, such as for
% the acknowledgment of grants, issue a \IEEEoverridecommandlockouts
% after \documentclass

% for over three affiliations, or if they all won't fit within the width
% of the page (and note that there is less available width in this regard for
% compsoc conferences compared to traditional conferences), use this
% alternative format:
% 
%\author{\IEEEauthorblockN{Michael Shell\IEEEauthorrefmark{1},
%Homer Simpson\IEEEauthorrefmark{2},
%James Kirk\IEEEauthorrefmark{3}, 
%Montgomery Scott\IEEEauthorrefmark{3} and
%Eldon Tyrell\IEEEauthorrefmark{4}}
%\IEEEauthorblockA{\IEEEauthorrefmark{1}School of Electrical and Computer Engineering\\
%Georgia Institute of Technology,
%Atlanta, Georgia 30332--0250\\ Email: see http://www.michaelshell.org/contact.html}
%\IEEEauthorblockA{\IEEEauthorrefmark{2}Twentieth Century Fox, Springfield, USA\\
%Email: homer@thesimpsons.com}
%\IEEEauthorblockA{\IEEEauthorrefmark{3}Starfleet Academy, San Francisco, California 96678-2391\\
%Telephone: (800) 555--1212, Fax: (888) 555--1212}
%\IEEEauthorblockA{\IEEEauthorrefmark{4}Tyrell Inc., 123 Replicant Street, Los Angeles, California 90210--4321}}

% use for special paper notices
%\IEEEspecialpapernotice{(Invited Paper)}

% make the title area
\maketitle

\newcommand{\red}[1]{{\color{red}[Note: #1]}}
\newcommand{\blue}[1]{{\color{blue}[Joseph: #1]}}

% As a general rule, do not put math, special symbols or citations
% in the abstract
\begin{abstract}
%\red{Abstract Under Development}
%The growing adoption of open-source code solutions and the emergence of transformer-based Large Language Models (LLMs) have profoundly impacted code development practices. Developers now have access to automated tools that swiftly resolve code-related issues. However, this ease of access has amplified security concerns in the realm of software development. Open-source solutions and code generation systems often prioritize functionality over security. Developers might unintentionally \color{magenta} obfuscate \color{black} existing code, creating potential vulnerabilities and inadvertently allowing malicious code to gain direct access to critical system functionality. Such incidents are further intensified due to the prevalent lack of security knowledge among developers. This work introduces an innovative solution based on Large Language Models and a new instruct-based dataset designed to identify code vulnerabilities in metamorphically generated code. Our solution also provides developers with mechanisms for repairing vulnerabilities and customized descriptions to deepen their understanding of the identified security weaknesses. As a practical application, we demonstrate the efficacy of our system in identifying, repairing, and describing zero-day and N-day vulnerabilities from the source code of IoT operating systems. Specifically, our system successfully identified 5 zero-day and 30 N-day vulnerabilities from the source code of IoT device operating systems. 
%\color{red} Our data and source code are publicly available at \color{black}.

Open Source Software (OSS) security and resilience are worldwide phenomena hampering economic and technological innovation. OSS vulnerabilities can cause unauthorized access, data breaches, network disruptions, and privacy violations, rendering any benefits worthless. While recent deep-learning techniques have shown great promise in identifying and localizing vulnerabilities in source code, it is unclear how effective these research techniques are from a usability perspective due to a lack of proper methodological analysis. Usually, these methods offload a developer's task of classifying and localizing vulnerable code; still, a reasonable study to measure the actual effectiveness of these systems to the end user has yet to be conducted. To address the challenge of proper developer training from the prior methods, we propose a system to link vulnerabilities to their root cause, thereby intuitively educating the developers to code more securely. Furthermore, we provide a comprehensive usability study to test the effectiveness of our system in fixing vulnerabilities and its capability to assist developers in writing more secure code. We demonstrate the effectiveness of our system by showing its efficacy in helping developers fix source code with vulnerabilities. Our study shows a 24\% improvement in code repair capabilities compared to previous methods. We also show that, when trained by our system, on average, approximately 9\% of the developers naturally tend to write more secure code with fewer vulnerabilities.

%Large transformer-based language models, such as BERT, have drastically altered the field of Natural Language Processing (NLP) in terms of code generation. However, they do not produce secure code and analysis of vulnerabilities that could help developers solve vulnerabilities faster. Therfore we propose an AI assisted tool to software developers analyze vulnerability with security analysis. We also provide a instruct based dataset that will train an LLM to generate secure code by determining the vulnerabilities.

%\red{Add one line-- IoT part here also as real life testing for 0 and n day}
\end{abstract}

% no keywords

% For peer review papers, you can put extra information on the cover
% page as needed:
% \ifCLASSOPTIONpeerreview
% \begin{center} \bfseries EDICS Category: 3-BBND \end{center}
% \fi
%
% For peerreview papers, this IEEEtran command inserts a page break and
% creates the second title. It will be ignored for other modes.
\IEEEpeerreviewmaketitle

\section{Introduction}
\label{1_introduction}

In the landscape of modern cyber security, Open-Source Software (OSS) has emerged as a critical component in a wide array of systems, ranging from IoT \cite{al2022idetect} platforms to essential software supply chains \cite{ladisa2023sok}. This prevalence positions OSS as a prime target for cyber adversaries. This is evidenced by high-profile breaches like the compromise of SolarWind's Orion platform \cite{peisert2021perspectives}, which affected approximately 18,000 stakeholders, including major government bodies and critical infrastructure providers. Such events not only highlight the vulnerabilities existing in widely-used software libraries \cite{log4j}, which can lead to extensive service disruptions \cite{synopsysrisc, log4j}, but also uncover the high impacts of cyber intrusions.

In response to these escalating security challenges, the interdisciplinary Open-Source Software Security Initiative (OS3I) was created, including agencies such as CISA, NSF, DARPA, and NIST. OS3I intends to develop software security guidelines and robust security measures \cite{fedreg}, which are increasingly important to consider in addressing cyber threats. Nevertheless, emerging technologies, including large language models such as GPT-4, escalate the complexity of ensuring the security of code generation. On the bright side, these technologies can streamline code development, yet they can also introduce significant security challenges \cite{khoury2023secure, sandoval2022security}, specifically by producing vulnerable code that is later used in production.

Given the dual nature of open-source code repositories or large language code generators, they require to be handled with caution. For novice developers, these resources can be invaluable as they facilitate faster progress and broader participation yet simultaneously introduce potential venues for adversaries to inject vulnerabilities into software \cite{democr1, democr2}. Current statistics indicate that nearly 40\% of open-source code resources may not meet stringent security standards \cite{pearce2022asleep} for a software application. This is a concerning trend as data from these resources is used to train large language code models. This scenario highlights the crucial role of developer proficiency in code security to mitigate these risks. Therefore, it is essential to recognize the challenges faced by developers, with some narratives framing them as potential "weakest link in the chain" \cite{green2016developers} in software security.  Current research emphasizes the need for robust security support and developer training \cite{green2016developers, acar2016you}. Surveys conducted by Assal et al. \cite{assal2019think} and Hermann et al. \cite{weir2020needs} echo this sentiment, revealing gaps in developers’ understanding of security practices and a lack of access to security experts. Proactively addressing these challenges becomes imperative to enhance the overall security posture of software development.

Recent advancements in automated vulnerability detection techniques \cite{zhou2019devign, li2018vuldeepecker, li2021vuldeelocator}, especially those based on transformer models \cite{nguyen2021regvd, islam2023unbiased}, have shown promise in identifying potential vulnerabilities in static source code. Transformer methods can learn patterns in source code to identify potential vulnerabilities. Further innovations are brought by graph convolutional networks (GCN) \cite{guo2020graphcodebert}, which can help security analysts determine specific vulnerability types. Another promising area is automated vulnerability repair (AVR) \cite{pearce2023examining, joshi2023repair, chen2022neural} systems powered by generative models \cite{2020t5, wang2021codet5, touvron2023llama, zelikman2023self}, which can autonomously generate recommendations for fixing vulnerable code. Automated vulnerability detection tools are essential for fending off cyber threats. While these tools are essential in mitigating cyber threats, they face challenges such as the diversity or complex business logic of source code, the evolving nature of cyber threats, the lack of comprehensive datasets, and the need for human expertise.

% Need to connect with prev. paragraph

Novel user-focused research suggests that software developers are in need of diagnostic information when addressing security challenges \cite{acar2016you, green2016developers}. While current vulnerability classification and localization methods play a critical role in discerning the specific location of code runtime failure, they cannot determine the root cause of the vulnerability. This problem is further exacerbated within large-scale applications, potentially comprising millions of lines of code, where crash reproduction can be exceptionally challenging due to environmental discrepancies of the system \cite{bellare1997forward, capobianco2019employing, han2020unicorn}. Although capturing program stack traces or logs \cite{ji2017rain, kang2011dta} might provide insights such as vulnerability-causing inputs or code break locations, the derived results are often incomplete as they do not necessarily identify the root cause of the vulnerability. Consequently, these run-time analyses from stack traces are incredibly lengthy; therefore, these require a long time to be sorted out, diminishing the usability of this method since it requires too much back and forth between code writing and full deployment. The lack of precise diagnostic information necessary to effectively rectify vulnerabilities highlights a pressing need for tools that not only identify but also explain the root causes of vulnerabilities, especially in large-scale applications with extensive code bases.

To address these challenges and provide developers with a more comprehensive understanding of vulnerabilities, we conducted an extensive survey with participants majoring in Computer Science from an R1 Research University. The insights from this survey revealed that existing automated vulnerability detection techniques often fail to provide developers with a clear understanding of a vulnerability's root cause as well as its classification and localization. Bridging this gap, we introduce T5-GCN, a novel approach for root cause diagnostics of software vulnerabilities. T5-GCN combines the power of a large language model (LLM) and a graph convolutional network (GCN) to classify and localize a vulnerability as well as identify the root cause, and provide developers with a static description in the source code. Additionally, T5-GCN integrates an explainability-based approach to facilitate extracting relationships between vulnerable statements and the pivotal contributors to the vulnerability's root cause. We conducted another human-centric analysis to evaluate our approach's efficacy and practical implications. The results demonstrate that T5-GCN can effectively aid developers in rectifying vulnerabilities and enhancing their proficiency in crafting more secure code. Moreover, T5-GCN demonstrated robustness in discerning the root causes of N-day and zero-day vulnerabilities. Overall, T5-GCN presents an innovative approach for root cause analysis of software vulnerabilities. 

%It has the potential to significantly contribute to the field of software security by enabling developers to rectify vulnerabilities and craft more secure code. 

The contributions delineated in this paper are as follows:
\begin{itemize}

    \item We present an introductory survey analysis to determine the flaws of the current systems and the shortcommings from the developers on security knowledge. Our primary analysis concludes that the root cause of the vulnerability is needed to help developers solve the vulnerability.

    \item We propose T5-GCN to find the root cause of a vulnerability using explainable techniques along with its classification and location of the existing vulnerability with a short static description of the vulnerability category. Furthermore, we performed an extensive evaluation with human subject participants to measure the effectiveness of our system compared to previous ones.

    \item We provide numerical analysis to show the quantitative results of our system. Furthermore, we also demonstrate the generalizability of our proposed approach by identifying N-day and 0-day vulnerabilities with their root cause in various open source applications.

\end{itemize}

Our data and source code are made publicly available here. \footnote{\url{https://anonymous.4open.science/r/Threat_Detection_Modeling-BB7B/README.md} }

\section{Background and Motivation}
\label{2_motivation}

As the complexity of software continues to increase \cite{alenezi2020relationship}, the significance of rigorous code security analysis in the post-deployment stage becomes paramount. In the absence of such measures, we leave our systems at risk, with potential vulnerabilities wide open to threat actors. These actors can exploit backdoors, which, although not deliberately created, may inadvertently remain within the codebase, posing a serious threat to system integrity and user trust. Proactively averting such vulnerabilities before deployment by analyzing static code testing constitutes a relatively less laborious approach, resulting in substantial savings in time and costs. Nevertheless, recent open-source static application security analysis tools exhibit limitations in generalizability. Furthermore, these tools do not offer actionable recommendations and comprehensive explanations to substantiate their decisions for the developers. Therefore, we deploy a survey to a targeted group of participants to further understand a software developer's current knowledge and needs to analyze a vulnerability and the usability of current SOTA techniques. The results from the survey will help us decide on the drawbacks of developer knowledge along with the challenges of current SOTA methods.

\begin{table}[ht]
\centering
\caption{Demography of the Participants from the Assisted and Controlled Group}
\begin{tabular}{lrrr}
\hline
\multicolumn{1}{l|}{}                      & \multicolumn{1}{l|}{\textbf{Control}} & \multicolumn{1}{l|}{\textbf{Assisted}} & \multicolumn{1}{l}{\textbf{Total}} \\ \hline
\multicolumn{4}{c}{\textbf{Undergraduate}}                                                                                                                               \\ \hline
\multicolumn{1}{l|}{Junior (Year 1 and 2)} & \multicolumn{1}{r|}{5}       & \multicolumn{1}{r|}{6}        & 11\\ \hline
\multicolumn{1}{l|}{Senior (Year 3 and 4)} & \multicolumn{1}{r|}{6}       & \multicolumn{1}{r|}{8}        & 14\\ \hline
\multicolumn{4}{c}{\textbf{Graduate}}                                                                                                                                    \\ \hline
\multicolumn{1}{l|}{MS}                    & \multicolumn{1}{r|}{6}       & \multicolumn{1}{r|}{6}        & 12\\ \hline
\multicolumn{1}{l|}{PhD}                   & \multicolumn{1}{r|}{8}       & \multicolumn{1}{r|}{11}       & 19                                                  \\ \hline
\multicolumn{4}{c}{Years of Programming Experience}                                                                                                             \\ \hline
\multicolumn{1}{l|}{0-2 Years}             & \multicolumn{1}{r|}{4}       & \multicolumn{1}{r|}{7}        & 11                                                  \\ \hline
\multicolumn{1}{l|}{2-5 Years}             & \multicolumn{1}{r|}{9}       & \multicolumn{1}{r|}{13}       & 22                                                  \\ \hline
\multicolumn{1}{l|}{More than 5 Years}     & \multicolumn{1}{r|}{8}       & \multicolumn{1}{r|}{9}        & 17                                                  \\ \hline
\multicolumn{4}{c}{Security Courses Taken}                                                                                                                      \\ \hline
\multicolumn{1}{l|}{Yes}                   & \multicolumn{1}{r|}{1}       & \multicolumn{1}{r|}{2}        & 3\\ \hline
\multicolumn{1}{l|}{No}                    & \multicolumn{1}{r|}{21}      & \multicolumn{1}{r|}{32}       & 53\\ \hline
\multicolumn{1}{l|}{Total}                 & \multicolumn{1}{r|}{22}      & \multicolumn{1}{r|}{34}       & $\leftarrow$ N = 56                        \\ \hline
\end{tabular}
\label{tab:demo_moti}
\end{table}

\paragraph{\textbf{Participant Demography}} In total, 56 participants completed our survey. We divided these participants into control and assisted groups. The control group only had access to the outputs from the SOTA models, and the assisted group had access to the outcome from our system only. At the end of our survey, we provided the participants with a demographic questionnaire, asking about their educational background, programming experience, and language preference. Table \ref{tab:demo_moti} shows the demographic information of our survey participants among two study groups, assisted and control.

The majority of participants are graduate students with CS majors with programming experience of more than two years. However, only one of the 23 participants from the controlled group had taken security-related courses, and two from the assisted group had taken security-related courses. While 5 and 6 participants from assisted and control groups were junior university students, and approximately 81\% were senior undergraduate or graduate (MS, PhD) students with more than two years of programming experience. Table \ref{tab:demo_moti} presents us that almost 60\% of the participant's control and 50\% of the assisted participants are graduate students, and approximately 70\% had more than two years of programming experience, making them a proper fit for our analysis. Furthermore, these groups represent potential students entering the software industry within a year or two or has some industry experience. Given their background and qualifications, they represent an ideal population capable of gaining proper insights into how the current education system has enriched their code security knowledge while simultaneously providing insights into the benefits and contributions of SOTA methods.

%Therefore, they represent the perfect mix of those who can provide proper insights into how the current education system has enriched their knowledge of code security and, at the same time, can determine how practical the existing SOTA tools are.

%\mj{One potential question reviewers will have is why this set of participants is representative of the problem we are studying? We need to add some sentences arguing why our participant population is representative of the code developers we are targeting for our solution.}

\subsection{Investigating Developers' Knowledge in\\ Security}
\label{moti_1}

%Software security plays a crucial role in ensuring the resilience of programs against various attacks. However, it is essential to acknowledge developers' challenges as they are often considered the weakest link in the chain regarding security \cite{green2016developers}. While developers are responsible for making final decisions on which code goes into production, recent research highlights the need for comprehensive security support and training for software developers to effectively address these issues \cite{green2016developers, acar2016you} before the code goes to production. One study by Assal et al. \cite{assal2019think} specifically examined the human factor in source code security. The survey highlighted the inadequate interaction between software developers and security experts.

\begin{figure}[ht]
    %\begin{framed}
        \centering
        
        \includegraphics[width=0.48\textwidth]{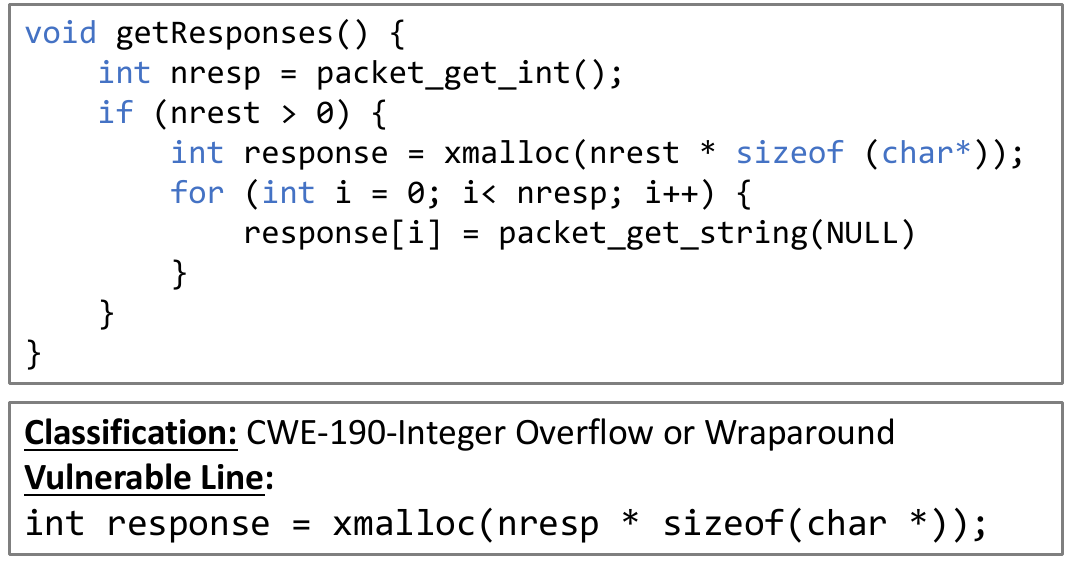}
        
    %\end{framed}
    
    \caption{Sample source code provided to the participants depicted at the top and output at the bottom provided by the SOTA methods. We conducted our initial survey by providing the participants with this information and determined their capability to repair vulnerability using these two outputs: classification and vulnerable line.}
    \label{fig:survey_sota}
\end{figure}

It is crucial to recognize and address the challenges developers face during code development. They are responsible for making final decisions on application design and implementation procedure, which ultimately reaches production. But a recent research underscores the necessity of providing comprehensive security support and training to empower software developers in effectively mitigating the security issues before code deployment \cite{green2016developers, acar2016you}. Another study by Assal et al. \cite{assal2019think} delved into the human aspect of source code security, shedding light on the insufficient collaboration between software developers and security experts.

\begin{table}[b]
\centering
\caption{Success rate of participants on completing ten different functions with proper security measures}
\begin{tabular}{l|r|r}

\hline
Function Name                  &   CWE Number &     Success     \\ \hline
\texttt{calculateCombinations} &       119    &      20\%       \\
\texttt{extractPrice}          &       264    &      40\%       \\
\texttt{exportPrices}          &       125    &      22\%       \\
\texttt{*loadPrices}           &       200    &      26\%       \\
\texttt{printMaxPrice}         &       416    &      46\%       \\
\texttt{validateUserCreation}  &       399    &      33\%       \\
\texttt{addUser}               &       20     &      18\%       \\
\texttt{removeUser}            &       476    &      26\%       \\
\texttt{promptUserCreation}    &       189    &      33\%       \\
\texttt{*resizeDatabase}       &       190    &      26\%       \\ \hline
\end{tabular}
\label{tab:moti_1}
\end{table}

%\paragraph{Insufficient Knowledge on Code Security:} 
Initially, we aimed to measure the percentage of developers who could write secure and vulnerability-free code without prior support. Therefore, in our initial investigation, we carefully crafted ten skeleton functions for a grocery store shopping cart and provided these to both groups. We also intentionally designed the function definition so that the response by the participants may fall susceptible to one particular type of vulnerability, as shown in Table \ref{tab:moti_1}. During this study, we did not provide any information regarding code security or tell them to be cautious about it. Our goal was to observe the natural coding style and quality of the participants and set a baseline to determine whether their natural coding practice included proper security measures. Details of the given task are explained in Section \ref{5_survey}.

Table \ref{tab:moti_1} shows the results from the assisted and control groups. From this table, we can see that out of the 56 participants who participated in our initial survey, CWE-119 and CWE-20 have one of the lowest success rates, 20\%, and 18\%, respectively. The success rate on the rest of the CWEs is slightly higher and peaks at 46\% for CWE 416. The vulnerabilities with the lowest success rate are about writing something out of the memory buffer or performing improper input validation. This relatively poor performance in writing correct and secure code also indicates the developers' indifference to safeguarding some of the variables they are supposed to operate.

%\mj{So what does this tell us? Write a couple of summary statements based on these results. Also, why CWE-119 and CWE-20 had higher success compared to others?}

%for all the cases, the number of vulnerable functions prevails compared to the non-vulnerable functions they have written. Each function is tied to a possible CWE category that we explained in detail in the Appendix \ref{}. Furthermore, we also can see that for the extractPrice function, we see the highest number of vulnerable function written by the participants, while exportPrice and printMaxPrice produces the maximum number of non-vulnerable code.
\color{black}

\subsection{Investigating Usability of Current SOTA Techniques}
\label{moti_2}

The usability of current publicly available code security tools is yet to be analyzed systematically. There are various popular static analysis tools such as Infer \cite{infer} and Cppcheck \cite{cppcheck} or deep learning methods to detect \cite{zhou2019devign} \cite{islam2023unbiased}, localize \cite{fu2022linevul} \cite{pornprasit2022deeplinedp} \cite{mirskyvulchecker} and repair \cite{fu2022vulrepair} \cite{pearce2023examining}, which can be used to address the security challenges faced by developers. While these tools and techniques provide some information to classify and localize vulnerability, they are not designed to act as an assistive/interactive tool for humans to understand the vulnerability (embedded in the code) in general or provide solid reasoning for their decision. The ideal goal of such tools should not be to offload a developer's task completely but to act as an assistive tool that would educate developers with appropriate security knowledge to help them become more self-sufficient in writing secure code.
%These studies further conclude that localizing vulnerability in code using AI-based methods does not provide proper training or understanding of the vulnerability to a developer.

\begin{table}[t]
\centering
\caption{The success rate of participants from the control group who could repair the function only when the vulnerable line and the CWE class of the vulnerability for each function were provided.}
\begin{tabular}{l|r|r}

\hline
Function Name               &      CWE-Number     & Success \\ \hline
\texttt{getValueFromList}   &      125        &      47\%      \\
\texttt{*callHelper}        &      416        &      48\%      \\
\texttt{SQLConnect}         &      264        &      20\%      \\
\texttt{readFile}           &      416        &      20\%      \\
\texttt{*createBoard}       &      20         &      46\%      \\ \hline
\end{tabular}
\label{tab:moti_2}
\end{table}

In order to understand how the SOTA methods help developers repair software vulnerabilities, we performed another investigation on the effect of the outcomes produced by current SOTA methods like VELVET \cite{ding2021velvet}, VulChecker \cite{mirskyvulchecker}, and LineVul \cite{fu2022linevul} which identify the vulnerability class along with the line of the vulnerability. In this study, we provide the control group of participants with five new vulnerable codes, the vulnerable line in that code, and the CWE classification of the vulnerability, similar to those provided by VELVET, VulChecker, and LineVul. Then, based on the information, we ask the developers to fix the vulnerability in the given code. We aim to determine the percentage of the participants that successfully repaired the vulnerable code. Figure \ref{fig:survey_sota} shows a sample code and SOTA outputs provided to the control participants to solve the vulnerability.

%\color{red}
%Comment
%\color{black}

For each of the five vulnerable functions, we generate outcomes for each of them from all the current SOTA methods. Our security experts then analyzed the outcomes for each function and determined the correct output. We do this to ensure that none of the outcomes provided to the participants were false positives or false negatives. After ensuring the correctness of all outcomes, we provided the participants with the five functions and the verified outcomes. Table \ref{tab:moti_2} shows that a majority of developers could not fix the given vulnerable code with the provided information. While the classification and localization provided some amount of insight on the code vulnerability to the developers, from the generated outcome by the developers, we can safely assume that these methods helped to solve less than 50\% of the vulnerability. While this is an improvement compared to writing code without any suggestions, if we consider a real-world situation, we see a possibility that approximately 50\% of code in production is still susceptible to vulnerability.

%\textbf{Interview}

% Drawing Conclusive Insights

\subsection{Concluding the Findings}
%(Recheck again: Backtracking towards solution)

While we investigated the challenges of current SOTA techniques to identify the benefits received by the developer, we asked them some follow-up questions to find the gap to assist the developer in writing secure code. We provided the developers with structured questions followed by an open-ended interview to determine what type of system they expected from which they would benefit the most. Since only the control group has seen outcomes from the existing SOTA methods, we asked them some follow-up questions.

In the first question, we wanted to determine the number of participants familiar with CWE vulnerability categories and found that only 17\% of participants are familiar. In the following question, we decided to investigate whether they looked for the definitions or the descriptions of the CWE on the internet, and 26\% of the participants did not. Furthermore, approximately 68\% mentioned that the description would help them greatly instead of looking it online. The last question we asked the participants was open-ended. Here, we asked the developers why they could not solve the vulnerability given the information. Almost 62\% of the developers responded that they did not know which line to look for to fix the particular vulnerability. Furthermore, 66\% of the developers also mentioned that if they were given the line or the root cause line they have to edit to repair along with the description, it would benefit them heavily.

Our primary analysis demonstrates the lack of developer training on code security. Further, we highlighted the failure of current SOTA techniques in assisting developers in repairing vulnerabilities. From the open-ended interview with the developers, we concluded that if developers are given a system that produces the root cause of the vulnerability with a reasonable description of the vulnerability, it would benefit them greatly.

\color{black}
\color{black}

%\section{Threat Model}
%\label{threat_model}
%\input{sections/threat_model}

%\section{Vulnerability Explanation Definition}
\section{Preliminaries}
\label{4_multitask_vulnerability}
\subsection{Threat Model} 

In exploring our threat model, we assume that rule or AI-based tools offload the task of vulnerability repair from humans. Therefore, we consider two types of threat actors in the software development landscape. The first category is the human developers who write vulnerable code because of a lack of proper security training on code. The second threat actor category is open-source platforms like ChatGPT, GitHub, or StackOverflow, where the same developers go and reuse code without security concerns.

%These tools do not act as a collaborator with human developers; instead, they provide static information that may help senior developers, but this information will hardly be helpful for young software developers. Furthermore, since these tools are not generalizable enough with high false positives and negatives, they provide incorrect information to developers, giving them a false sense of security.

Our threat model mainly focuses on the vulnerable code developers write in C/C++. Our primary analysis shows that software developers are primarily vulnerable to addressing null pointer or resource management-related vulnerabilities. The vulnerabilities at these levels could compromise the physical memory and CPU caches by gaining privileged root mode access. The attackers could exploit such vulnerabilities through various attack scenarios, like buffer overflow, code injection, improper operations within a memory buffer, or similar vulnerabilities related to these. These vulnerabilities can subsequently provide control over the system, enable data theft, or even launch further attacks. Our approach to minimizing vulnerability in source code is twofold. We propose a system to aid developers in finding the root cause of vulnerability along with the CWE category, complemented by a description of the vulnerability and the vulnerable line. Secondly, with the combination of these pieces of information, we aim to provide usable security to aid developers in writing better and more secure code with the assistance of AI.

%\color{red}(Consider illustrating this with a flowchart or diagram for visual representation)
%\color{black}.
\subsection{Problem Formulation}

%\mj{This subsection starts a bit wierdly. You should typically start with the problem we are trying to solve.} 

In order to assist developers in fixing a vulnerable code and making them aware of introducing future vulnerabilities, in this study, we want to address the problem of finding the root cause of vulnerability along with its CWE classification with a short static description and the vulnerable line. To prepare the input for the GCN, we convert each program function, denoted as $p_i$, into a multi-edged graph, $\mathcal{G}_{raph}$. This graph is constructed such that the set of nodes, $T$, represent the programming language tokens, and the edges, $E_{dge}$, indicate the connections between these nodes. We transform the edge pairs into an adjacency matrix $A$.

Our methodology employs a multitask function that identifies the root cause of vulnerability with its classification and localization. Furthermore, we provide the developers with a static description curated explicitly for novice developers. Initially, it ascertains the vulnerability category, denoted as $CWE$, that a vulnerable input function $p_i$ aligns with. After classifying $p_i$, the model further localizes the starting ($L_{start}$) and the ending ($L_{end}$) vulnerable lines, $[L_{start}, L_{end}]$. In the context of our system, we define three outputs: the vulnerability classification $CWE$ with a static description, the vulnerability localization range $[L_{start}, L_{end}]$, and the root cause of vulnerability $V_{root}$.

\textbf{Root Cause Analysis through Explainability:}
The proposed architecture uses explainable artificial intelligence techniques to discern the impacts contributed by each program token, denoted $t_i \in T$, on the vulnerability prediction of a given program.

This model, initially trained to classify and localize the vulnerabilities, garners a holistic understanding of the vulnerabilities and correspondingly attributes weights to each token in the source code. Beyond merely localizing the vulnerable lines, our model proficiently comprehends individual tokens' implications. By prioritizing specific tokens over others, the model inherently puts more significance to the critical tokens that show a higher likelihood of engendering vulnerability.

For the scope of this research, the weight associated with a token $t_i$ is symbolized as $\phi\textsubscript{i}$. The predictive output delivered by our model for one program is mathematically defined as follows:

\begin{equation}
\hat{A} = f({T}) = \phi \textsubscript{0} + \sum_{i=1}^{N} \phi \textsubscript{i} {t_i}
\end{equation}

Where $\hat{A}$ is the set of attributions for each token in $T$, and $N$ is the total number of tokens. In this expression, the weight $\phi\textsubscript{i}$ encapsulates the contribution of a token $t_i$ to the model's overall output.

\textbf{Assessing System Efficacy:}
To address the issues we found in our initial analysis in Section \ref{2_motivation}, we try to address them by answering the following three Research Questions (RQs):

\noindent\textbf{RQ1:} \textit{Using the root cause of vulnerability, how effectively is our system assisting software developers in fixing code and educating developers in writing secure code with fewer vulnerabilities? }

\noindent To find the root cause of source code vulnerability, we used an explainability-based technique to determine the importance of tokens. Then we propose an in-depth survey analysis to determine how effective our system is compared to the current SOTA methods.

\noindent\textbf{RQ2:} \textit{How efficiently can we classify and localize the vulnerability in addition to finding the root cause vulnerability of source codes?}

\noindent To find how efficiently we can localize vulnerability, we use a metric called IoU. Furthermore, in order to measure the classification accuracy, we measure F1 Accuracy scores. The higher classification and localization performance will ensure the root cause analysis's effectiveness.

\noindent\textbf{RQ3:} \textit{Does our root cause detection system generalize enough to identify zero- and $n$-day vulnerabilities from the wild?}

\noindent To answer this question, we analyzed several open-source projects written in C/C++, and our internal security experts manually checked the validity of the root cause provided by our system.

%<Intuitive approach: To answer each RQ a paragraph explaining the workflow>

These RQs serve as a benchmark to evaluate our system's vulnerability detection and localization capabilities, its competency in providing actionable insights for vulnerability remediation, and its adaptability in discerning new vulnerabilities across diverse environments. Subsections \ref{rq1}, \ref{rq2} and \ref{rq3} provide in-depth detail in our attempt to answer the above three research questions.

\section{User-Focused Survey Design}
% (If possible find a better name)
\label{5_survey}
%\subsection{User-Focused Study Design (Survey Design Goes Methodology)}

\begin{figure*}[t]
    \centering
    \includegraphics[scale=0.70]{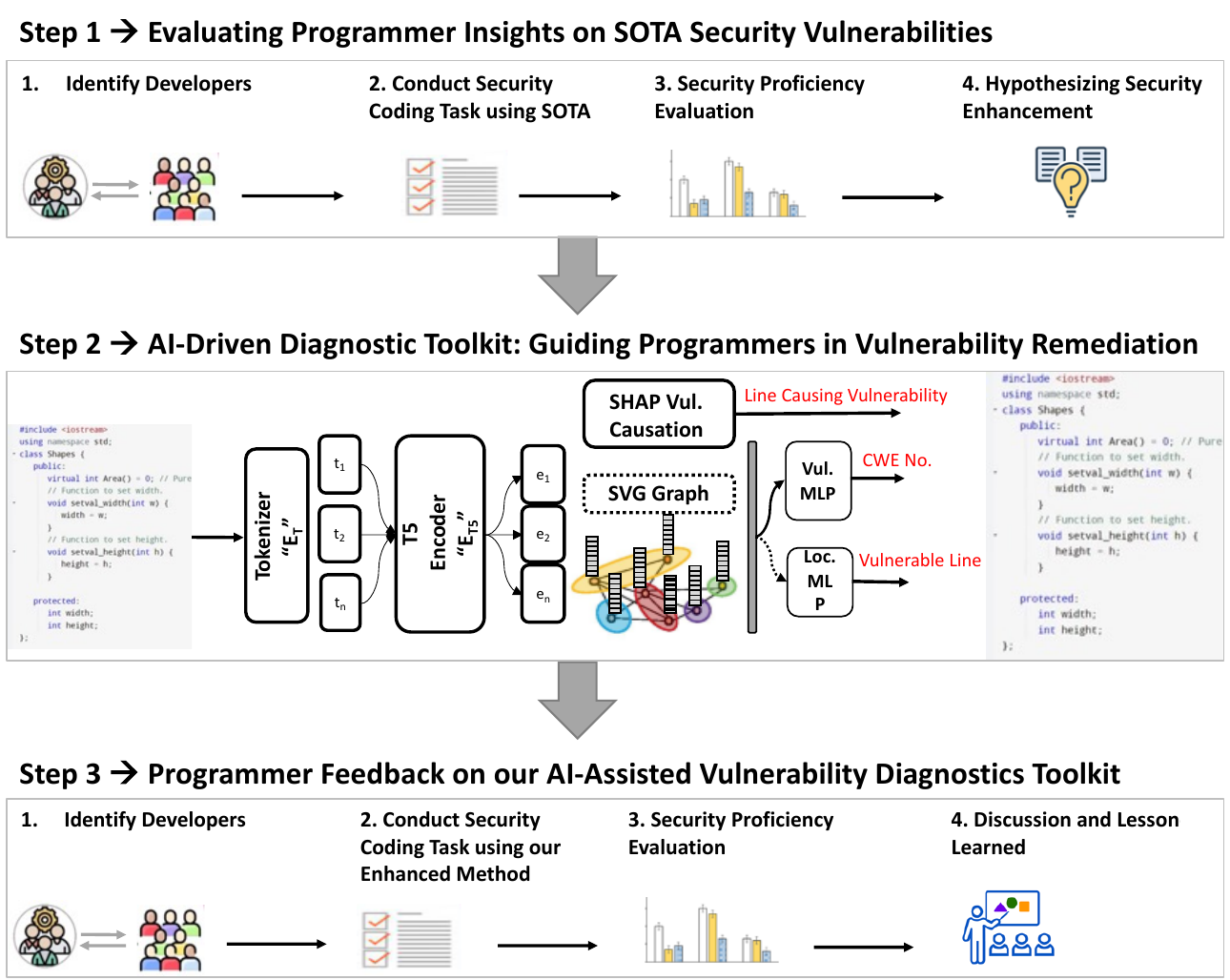}
    %\vspace{-0.3cm}
    \caption{Our proposed approach is organized into three pivotal steps (1) Evaluating programmer insights on current state-of-the-art security vulnerabilities; (2) Introducing an LLM-powered diagnostic tool that assists programmers in vulnerability remediation; and (3) Analyzing programmer feedback on the toolkit's effectiveness.}
    \label{fig:fig_2}
    %\vspace{-3mm}
\end{figure*}

We designed a user-centric survey to evaluate the effectiveness of our proposed T5-GCN for addressing code vulnerability issues and aiding developers in learning secure coding practices with less reliance on AI tools. Figure \ref{fig:fig_2} shows the entire workflow and architecture of our study and step 1 and 3 in the figure demonstrates the survey analysis presented in this section.

%While our proposed system does not provide any code completions, we provided information to the developers to help them fix any possible vulnerability in their code. Our approach equips developers with insights enabling them to address code vulnerabilities, ensuring they maintain complete control over their code and its underlying logic. We tested our proposed system using undergraduate and graduate students from Computer Science with programming experience from an R1 research university. 

\subsection{Participant Recruitment}

We recruited our participants from the Computer Science department, enlisting both undergraduate and graduate students who have either completed or are currently enrolled in at least one programming language course. We selected participants with a wide range of educational backgrounds: 1) Undergraduate Students with programming experience, 2) Graduate Students without industry experience, 3) Graduate Students with industry experience, and 4) Graduate Students with research experience.

The recruitment process took place via email invitation and flier distribution, with the approval and support of the department's admin. Furthermore, since many of these participants will join the industry after graduation, their selection is strategic, considering they likely possess limited knowledge and training in code security. The selected group allows us to estimate the effect of insufficient code security knowledge among emerging professionals. Our recruitment process emphasized evaluating their code-writing quality, and therefore, we deliberately omitted the focus on code security practices in our survey. This omission aimed to prevent potential bias, as participants could have otherwise prepared in advance, skewing the results. 

\paragraph{\textbf{Compensation and Approval}}Each participant received a compensation of US\$25 for their contribution to the study. This study involved human participants and was approved by the institution’s Institutional Review Board (IRB).

%The study commenced with participants tackling programming tasks, each embedded with potential Common Weakness Enumeration (CWE) vulnerabilities. These tasks were designed in the C/C++ programming language to ensure consistency (what consistency?) in evaluation. Furthermore, we chose C/C++ as the programming language for our survey since many memory and resource-related vulnerabilities directly arise from it \cite{msrc}.

For a broader understanding of the impact of our proposed system, participants were randomly allocated to two distinct groups: the Control Group and the Assisted Group. Participants were provided with four types of information based on which group they were assigned to: 1) the specific vulnerable line(s) of code, 2) the vulnerability category, 3) the description of the category, and 4) the root cause of the vulnerability. The control group was provided with the vulnerable line and its associated category. In contrast, the assisted group was provided with the root cause of the vulnerability, along with classification, description and localization.

%The control group imitates the current state-of-the-art methods, including assisting the developers by providing localization as the highest level of information. However, the assisted group imitates our proposed system, where the developers are provided with the root cause of the vulnerability and a description that helps them understand the vulnerability in the code. We hypothesize that a clear comprehension of the vulnerabilities, supported by their classifications and detailed descriptions, will empower developers to amend the code adeptly, obviating the need for additional resources.

\subsection{Survey Design}

Our survey is segmented into two stages to rigorously evaluate our model's effectiveness and usability, namely 1) The baseline and 2) the evaluation stages.

%\textbf{Learning Stage:} 
%Randomly assigned to the controlled or assisted groups, participants tackled 10 code vulnerability problems. The controlled group received vulnerability localization and classification, while the assisted group additionally gained insights into the root causes. We assume this division will cause some participants to learn and generalize code vulnerability better. After showing them the training codes, we provide the users with some exercise problems to understand how efficiently they are fixing vulnerabilities.

\paragraph{\textbf{Baseline Stage}} In the baseline stage, we asked participants from assisted and control groups to complete ten C functions associated with a grocery store management system. The participants have to download C source files containing skeletal code (i.e., a function declaration without implementation) for each function, accompanied by implementation instructions, parameter and return value details, and example usages. Participants were required to document the start and completion times for each function. Listing \ref{lst:listing-cpp} illustrates one of the ten functions the participants need to implement. The tasks involved core C programming concepts like file I/O and memory management. However, we ensured that the implementations of these functions steer clear of advanced data structures to avoid unnecessary complexity. Essential C libraries and user-defined structures were provided with the skeleton code beforehand. Participants were neither encouraged nor discouraged about using and utilizing resources from online or any code-generative AI, and they could use any code editor or IDE based on their personal preferences. The baseline stage is presented in Subsection \ref{moti_1}

%Participants could consult online resources and utilize any text editor or IDE based on their preferences. 

\begin{lstlisting}[style=CStyle, label={lst:listing-cpp}, caption=A function that primarily checks if the participant handled integer overflow and properly set up a base case to avoid a stack overflow if solved recursively.]
long calculateCombinations(long numItems)
{
    /**
     * Problem:
     *  In a grocery store, there are different types of items available for purchase. The store
     * manager wants to calculate the total number of possible combinations of items that a
     * customer can buy. This function should return the factorial of numItems (i.e. numItems!)
     *
     * Example:
     *  calculateCombinations(3) returns 6 (3! = 3 * 2 * 1 = 6)
     *
     * Parameters:
     *  numItems: the total number of items in the store.
     *
     * Returns:
     *  the total number of possible combinations.
     */
    //  --> PARTICIPANT: ENTER HERE TIME STARTED <--
    return 0;
}
//  --> PARTICIPANT: ENTER HERE TIME COMPLETED <--
\end{lstlisting}

Furthermore, we provided the control group with five functions intentionally associated with a particular CWE number. This means that if the participant makes any security mistake, it will be tied to that CWE number. We provided the two outcomes provided by the SOTA techniques. They were the vulnerable line and the CWE category of the vulnerability. The task was to resolve vulnerability with the given information. We conducted this part of the survey in Subsection \ref{moti_2}.

\paragraph{\textbf{Evaluation Stage}}
In this stage, we assess the participants' ability to repair code and extend the knowledge they have acquired regarding the rectification of security-related code vulnerabilities. The assessment revolves around the five functions we previously provided to the control group in subsection \ref{moti_2}. However, this time, the outcomes we provided to the assisted group were the vulnerable line, the CWE class with description, and the root cause of vulnerability. We asked the assisted group to repair the given five codes. Furthermore, we asked the developers from both groups to rewrite the ten functions they wrote in the baseline stage in Subsection \ref{moti_1} to fix any possible security issues. The goal was to determine how developers from both groups improved their knowledge using SOTA methods in the control group compared to our technique in the assisted group. This part was done in Subsection \ref{rq1}.

We described the workflow of our survey in Appendix \ref{9_appendix}.

\subsection{Code Assessment}
We employed a dual-method evaluation approach to assess the participants' code regarding security and functionality. Furthermore, we categorized identified bugs according to their CWE classification \cite{cwe}.

\paragraph{\textbf{Run-time Analysis}} For the run-time of the code written by the participants, we generated two sets of test cases for each function—the first set aimed at evaluating the functionality, ensuring that the code meets the intended requirements. The second, more intricate set was designed to unveil potential vulnerabilities, with specific inputs crafted to induce run-time errors if vulnerabilities were present. A function passing all test cases, functionality, and security was deemed free of vulnerabilities.

\paragraph{\textbf{Manual Analysis}} While most participants submitted a fully working code, some provided partially written non-compilable code. Furthermore, some participants also use pseudocode as a solution to vulnerability. Therefore, we also do a manual code review by our internal security experts to identify potential vulnerabilities and fixes. Our security experts initially manually checked the functionality by reading code and try to estimate the outcome based on some input test cases. Furthermore, for security testing, our security experts check whether the function would fall into a runtime error in the case of NULL, negative, or overloading numerical limits in the code. Based on the partially written code or pseudocode, if the function is possibly breaking at these inputs, the security experts deem it as a vulnerable function.

\section{Proposed System Architecture}

% (Survey and Methodology are combined methodology, Arc fig go before 5)LLM Powered Diag. ToolKit
\label{6_methodology}

We provide an end-to-end system to analyze source code vulnerabilities and demonstrate our system's capability to assist developers in writing secure and vulnerability-free code. Figure \ref{fig:fig_2} depicts the overall architecture of our proposed vulnerability resolution and evaluation procedure. Step 2 in this figure explains the methodology for code vulnerability classification, repair, and finding the root cause of the vulnerability.

% SVGCodeT5
%\subsection{Root Cause of Vulnerability}
\subsection{Code Vulnerability Detection and\\ Classification}
%1.Title update to causation, 
%2. But keep the order as before

%\color{black}
For code vulnerability detection and classification, the input source code has to go through pre-processing steps and, finally, through our proposed T5-GCN.

\paragraph{\textbf{Source Code Representation}} In this pre-processing step, the input is an entire function of source code, which may be vulnerable or non-vulnerable. We initially employ the CodeT5 \cite{wang2021codet5} tokenizer, which tokenized words using a byte-pair fashion \cite{sennrich-etal-2016-neural}. CodeT5 tokenizer was pretrained in programming languages like C/C++ to extract the set of tokens $T$ from a given function $p_i$.

We analyzed the individual functions by random sampling from our datasets and found that the average number of tokens is approximately 490. Therefore, we propose to select 512 as the maximum number of tokens, and we trim the length of the set of tokens $T$ to 512. Moreover, we add two unique tokens, $<BOS>$ and $<EOS>$, at the beginning and end of the program as a separator. If the length of the program is less than 512, we use a unique token $<PAD>$ to resize the length to 512. After finalizing the nodes, we develop $\mathcal{G}$ by connecting the nodes using SVG \cite{islam2023unbiased}.

\paragraph{\textbf{Source Code Semantic Graph Representation}} We have refined the process of root cause analysis within source code by enhancing the capabilities of Graph Convolutional Networks (GCN) through the incorporation of a Semantic Vulnerability Graph (SVG) \cite{islam2023unbiased}. The SVG combines four distinct categories of edges, encompassing data \cite{zhou2019devign}, control \cite{zhou2019devign}, sequential \cite{huang2019text, zhang2020every}, and poacher flow \cite{islam2023unbiased} relationships. These four edge categories comprehensively capture the source code's syntactic and semantic attributes.

By combining these diverse graph types, the GCN gains an intricate understanding of the source code, enabling a contextual interpretation and passive runtime understanding \cite{islam2023unbiased} of the static code. This contextualization facilitates the creation of relational representations for different program tokens, significantly enhancing the system's ability to pinpoint the tokens responsible for underlying vulnerabilities. This holistic approach to vulnerability analysis represents a significant advancement in source code security analysis.

\paragraph{\textbf{CodeT5 Encoder}}
Our proposed system is powered by CodeT5 \cite{wang2021codet5}, a large language model that adopts the encoder-decoder architecture inspired by T5 \cite{2020t5}. It effectively captures the syntactic structure of code and utilizes positional information associated with each token to facilitate token-based localization.

We use the encoder of CodeT5, which consists of multiple layers of self-attention and feed-forward neural networks, to generate the embedding of each node or token in our graph. The self-attention mechanism computes attention weights to capture the input sequence's interdependencies and relationships between elements. This allows for encoding contextual information from the nearby tokens in the code. The output from the self-attention layer is then passed through a feed-forward neural network, which applies a nonlinear transformation independently at each position and finally outputs an embedding vector $E$ of size 768 for each token. This embedding vector acts as the node representation for each token, which is then converted into an adjacency matrix using SVG \cite{islam2023unbiased} and passed to the GCN layer.

%\paragraph{\textbf{Ensemble VulCodeT5}}

\paragraph{\textbf{T5-GCN}} Graph Convolution Network (GCN) attempts to comprehend the correlation between any pair of node embeddings of tokens from code we got from the CodeT5 encoder. We introduce a two-layered GCN with a residual connection. Mathematically, we implemented GCN as follows:

\begin{equation}
\label{eqn:7}
F_{GCN} = H\textsuperscript{(n + 1)} = H\textsuperscript{n} + \sigma \biggl( W^n_{GCN} H\textsuperscript{n} A \biggr )
\end{equation}

Here, $W^n_{GCN}$ represents the learnable weights at the $n$-th layer, and $H\textsuperscript{n}$ is the feature representation of all tokens $T$ from a function $f_i$ at the $n$-th layer. $H\textsuperscript{(0)} = E$ and $A$ represents the adjacency matrix. The multiplication of the matrices $W^n_{GCN}$, $H_n$, and $A$ is followed by an activation function $sigma$ (e.g., $ReLU$). $F_{GCN}$ is the final representation generated by our proposed GCN.

%After the GCN layer, we add two dense layers in parallel, as depicted in Step 2 in Figure \ref{fig:fig_2}, one for CWE classification of the vulnerabilities and another for localizing the vulnerable line.

%%%%%% Ignoring Most of Loss Function Part
\paragraph{\textbf{Loss Function}} We use the Focal Loss function \cite{lin2017focal}, built on top of CrossEntropyLoss, which can handle possible data imbalance issues as identified by \cite{islam2023unbiased} for vulnerability classification purposes. Our Focal Loss function stands thus:

\begin{equation}
    \label{eqn:loss1}   
FocalLoss(p_{rob}^t) = -\alpha (1 - p_{rob}^t)^\delta \log(p_{rob}^t)
\end{equation}

In this instance, $\alpha$ denotes the balancing factor between the number of vulnerable and non-vulnerable code samples, while $p_{rob}^t$ is the probability distribution of our model's output. We use $\delta$ as an adjustable parameter that distinguishes between easy and hard examples \cite{lin2017focal}.

\paragraph{\textbf{Detection and Classification}}For vulnerability detection and classification purposes, we use the feature vector $F_{GCN}$, produced by our proposed T5-GCN. We added a dense layer after the feature vector layer generated by GCN. The dense layer \texttt{Vul. MLP} generates the CWE Number of the vulnerable code if a vulnerability exists, as depicted in Step 2 of Figure \ref{fig:fig_2}. If no vulnerability exists, the \textit{Vul. MLP} layer generates an output of 0. Furthermore, for each identified vulnerability classified by a CWE number, we provide a static description of the identified vulnerability.

\subsection{Identification of Vulnerable Lines}
In order to find the vulnerable lines, our proposed model identifies a block of code by generating the starting and ending lines of the vulnerable code. The second dense layer \texttt{Loc. MLP} generates $L_{start}$ and $L_{end}$, the line range where the vulnerability exists. Therefore, we connect $F_{GCN}$ with another dense layer \textit{Loc. MLP} for finding the vulnerable line. 

Since line numbers vary depending on the position of the vulnerable line in code, we designed the identification of vulnerable lines as a regression problem. Hence, we apply Mean Squared Error (MSE) loss for vulnerability localization. Our MSE loss function is defined as follows:

\begin{equation}
    \label{eqn:loss2} 
MSE = \frac{1}{n} \sum_{}^{} (L - \hat{L})^2
\end{equation}

where $L$ is the original outcome and $\hat{L}$ is the outcome from the model.

\subsection{Root Cause of Vulnerability}

After the model is sufficiently trained to classify and localize the vulnerability, we find the root cause of the vulnerability using our trained model. We employed DeepLiftSHAP attribution scores. We hypothesize that, since the model can effectively classify and localize the vulnerability,  we determine to utilize the model's understanding of vulnerability to determine the contribution of each token to find the root cause of vulnerability.

DeefLiftSHAP is an explainability technique for neural networks based on executing a SHAPly \cite{lundberg2017unified} variant of the original DeepLift \cite{shrikumar2017learning}. Combining DeepLIFT and SHAPly, DeepLiftSHAP operates on deep learning frameworks to explain neural network models. We generate attribution scores based on the DeepLiftShap \cite{lundberg2017unified}, where we generate the code token attribution scores based on our proposed Algorithm \ref{alg:algorithm}. We sum up the attribution scores of each token in a line to generate an attribution score for each line. Here, $\hat{A}$ is the set of attribution scores for all tokens in $T$ of a function $p_i$, where, $\hat{A} \in \{a_1, a_2, ... , a_m \}$. After generating scores for each line or statement, we consider the line with the highest attribution values before $V_{Start}$ as the root cause of the vulnerability. 

%Algorithm \ref{alg:algorithm} to find the root cause of vulnerability. $L_{vuln}[L_{Start}, L_{End}]$ is the localized vulnerable line, and $c$ is the top $c$ statements with the highest attribution scores that we want to consider as the root cause of the vulnerability. The function $getBenignLines$ gets all the lines before the vulnerable line $L_{Start}$ starts.

%\algnewcommand{\LeftComment}[1]{\Statex \(\triangleright\) #1}

% // Step 3: Apply DeepLIFT
%     Initialize DeepLIFTContributions as an empty list
%     For each instance x in ProcessedX:
%         contributions = DeepLIFT(M, x, R)
%         Add contributions to DeepLIFTContributions

\begin{algorithm}[t]
    \caption{Token Attribution for Root Cause Vulnerability}
    \label{alg:algorithm}
    \textbf{Input}: $model$,  input program $p_i$ \\
    %\textbf{Parameter}: Optional list of parameters\\
    \textbf{Output}: Explainable attribution scores $\hat{A}$ for tokens $T$  
    %\Procedure{\textit{VULNERABILITY\_EDGES}}{$T$}

    \begin{algorithmic}[1] %[1] enables line numbers
        \STATE $T$ = $Tokenizer (p_i)$
        
        \STATE $\hat{A}$ = [$t_i$: 0 for $t_i$ in $T$] \\
        
        %\COMMENT{Calculating Reference Output from Model} 
        \STATE $original_{pred}$ = $model(T)$

        \STATE $contrib_{subset}$ = $[]$
        %\COMMENT{DeepLift Contributions}
        \FOR{each $t_i$ in T}
            \STATE $contribution$ = DeepLIFT($t_i$)
            \STATE $contrib_{subset}.append(contributions)$

        \ENDFOR
        
        %\STATE $S$ = $Subset(T)$

        \FOR{Each $s_{sub}$ in $contrib_{subset}$}
            
            \STATE $subset_{pred}$ = $model(s_{sub})$

            \STATE $marginal_{contr}$ = $original_{pred}$ - $subset_{pred}$

            \STATE $norm_{contr}$ = $marginal_{contr}$ / $len(s_{sub})$

            \STATE $\hat{A}.append(norm_{contr})$
        \ENDFOR

        \STATE \textbf{return} $\hat{A}$
    
    \end{algorithmic}
\end{algorithm}

\section{Experiments and Discussions}
\label{7_experiments_discussion}

We aim to provide an LLM-powered code assistant to help developers write more secure code and potentially educate developers in writing vulnerability-free code. Therefore, in our experiments, we initially measure our system's qualitative and quantitative results, followed by an analysis of its generalizability. 

%We analyze the capability of our system by answering the 3 RQs we mentioned in Section \ref{4_multitask_vulnerability}.

\subsection{Experimental Datasets}
\label{sec:5_dataset}

\paragraph{\textbf{D2A and BigVul}} Since we provide an end-to-end solution for vulnerability analysis, we only focus on the datasets containing source code from real-world applications. We use Big-Vul \cite{fan2020ac} and D2A \cite{zheng2021d2a} for vulnerability classification, localization, and root cause analysis. BigVul provides ten vulnerability categories that fall within the top 25 CWE vulnerabilities mentioned at CWE \cite{cwe}.  D2A contains open-source projects from GitHub, and labels were created using commit filtering and static analyzer tools. The BigVul dataset provides CWE numbers of vulnerabilities, while D2A does not provide a vulnerability class, so we use it only for vulnerability detection and localization.

\paragraph{\textbf{IoT OS Repositories}}
We collected a dataset from six OS repositories from GitHub to test the capability of our system in a real-world operational analysis for wild N-day and zero-day program samples. First, we downloaded six IoT operating repositories: TinyOS, Contiki, Zephyr, FreeRTOS, RIOT-OS, and Raspberry Pi OS. Next, we scanned the entire repository of these operating systems using JOERN \cite{joern}, a tool specially designed to monitor and analyze large repositories. Then, we used JOERN command line interface to split the C/C++ files into functions for operational analysis.

\subsection{Evaluation Metrics}

We analyze the qualitative result of our root cause analysis from our survey. However, for quantitative analysis, we use some standard metrics. We achieve vulnerability localization by establishing a boundary between vulnerable lines (starting and ending vulnerable lines). Therefore, we employ the Intersection of Union (IoU) as our evaluation metric. Since our input data consists of source code, which is linear single-dimensional data, unlike a 2D image, we modify the IoU formula \cite{rezatofighi2019generalized} to 1D scale. Let's consider our model predicts the localization line boundaries from $ \hat{Vul_{Code}} = [\hat{L_{Start}} - \hat{L_{End}}]$ and $Start <= End$ and the ground truth for localization is $Vul_{Code} = (L_{Start}, L_{End})$. So the IoU is:

\begin{equation}
\label{eqn:11}    
IoU = \frac{|\hat{Vul_{Code}} \cap Vul_{Code}|}{|\hat{Vul_{Code}} \cup Vul_{Code}|}
\end{equation}

If the value of IoU is zero, Equation \ref{eqn:11} demonstrates that $\hat{L}$ and $L$ do not overlap, indicating that the model cannot accurately localize a single vulnerable line. If the value is 1, the model can accurately anticipate all vulnerable lines. However, we used the standard metrics including accuracy, precision, recall, and F1 score metrics for vulnerability classification purposes.

%\textbf{RQ1: Based on our classifier, can it learn to identify and localize vulnerabilities in real-world source code?}

\subsection{Experiments}

We randomly divided our datasets into 80:10:10 ratios for training, validation, and testing during the experiment. We utilized a 12-layer CodeT5 encoder to generate the embeddings and a two-layer GCN with a residual connection to generate feature vectors for each function $p_i$. The final feature representation vector produced by GCN is of size 512. We trained the model for 20 epochs with a maximum token length of 512 for each function using a learning rate 6e-6. We used eight A100 NVIDIA GPUs for our experiments.

\paragraph{\textbf{Multitask Training}} As depicted in Figure \ref{fig:fig_2}  during the training stage, we perform classification and localization for vulnerability analysis. We use cross-entropy loss for vulnerability detection and MSE loss for localization purposes. Furthermore, we provide a static description of the class of vulnerability to the developers.

Since we are identifying ten vulnerability classes, there are ten layers of neurons for vulnerability classification. However, we also provide localizing vulnerability by providing the line range of the statements where the vulnerability exists. Therefore, we put two neurons for vulnerability localization to generate the line numbers for the first and last lines of the vulnerable statements. Finally, for root cause analysis, we used the trained model and utilized explainability to the root cause of the vulnerability.

In the remaining part of this section, we will conduct the experiments by answering the three research questions we defined in Section \ref{4_multitask_vulnerability}.

\begin{figure}[t]
    %\begin{framed}
        \centering
        
        \includegraphics[width=0.48\textwidth]{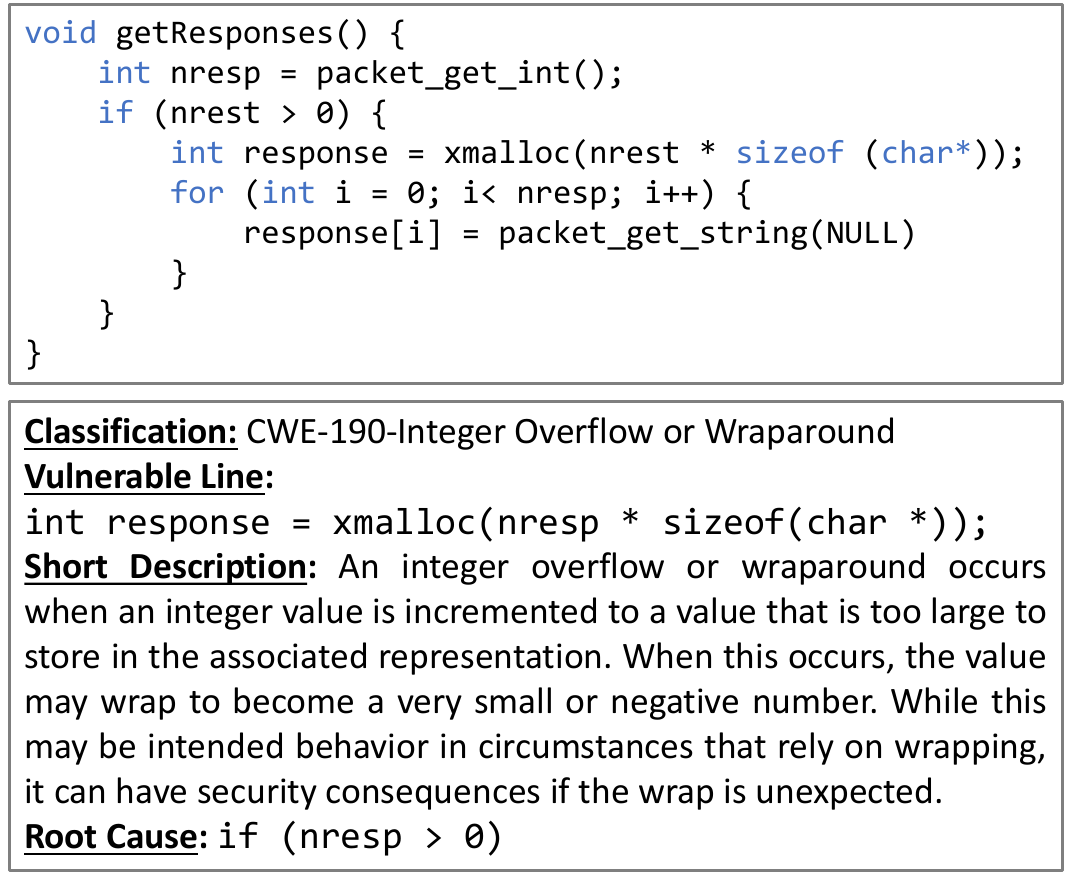}
        
    %\end{framed}
    
    \caption{Input code is depicted at the top and output produced our proposed system is depicted at the bottom. Input is the vulnerable code, and our system provides four types of output: 1) Classification, 2) Vulnerable Line, 3) Short Description, and 4) Root Cause.}
    \label{fig:survey_our}
\end{figure}

\subsection{RQ1: Developer Usability Evaluation}
\label{rq1}

%\textbf{Using the root cause of vulnerability, how effectively is our system assisting software developers in fixing code and educating developers in writing secure code with fewer vulnerabilities?}

\begin{table}[b]
\centering
\caption{The success rate improvement of the assisted participants compared to the controlled who were only given the vulnerable line and the CWE class of the vulnerability. Here the improvement depicts the percentage of improvement of the assisted group compared to the control group for vulnerability repair and how quickly the assisted group repaired vulnerability.}
\begin{tabular}{l|c|c|c}
\hline
\textbf{\begin{tabular}[c]{@{}c@{}}Function \\ Name\end{tabular}} & \multicolumn{1}{c|}{\textbf{\begin{tabular}[c]{@{}c@{}}CWE\\ Number\end{tabular}}} & \textbf{\begin{tabular}[c]{@{}c@{}}Repair\\ Impr.\end{tabular}} & \textbf{\begin{tabular}[c]{@{}c@{}}Time \\ Impr.\end{tabular}} \\ \hline
\texttt{getValueFromList} & 125             & 7\%                         & 10\%                      \\
\texttt{*callHelper}      & 416             & 15\%                        & 6\%                       \\
\texttt{SQLConnect}       & 264             & 11\%                        & 14\%                      \\
\texttt{readFile}         & 416             & 16\%                        & 9\%                       \\
\texttt{*createBoard}     & 20              & 24\%                        & 22\%                      \\ \hline
\end{tabular}
\label{tab:qual_com_1}
\end{table}

To qualitatively evaluate the usability of our system, we used the assisted group of participants to determine how our system can assist them in repairing code vulnerability. Furthermore, we also test the capability of our model to educate the developers in writing vulnerability-free code without assistance. To find the root cause of vulnerability, we use attribution scores generated using SHAPly. The top box in Figure \ref{fig:survey_our} shows the vulnerable code and the bottom box shows the root cause and all the other outcomes generated by our model and provided to the assisted group.

\paragraph{\textbf{Assisted Vulnerability Repair}}

Here, our goal is to determine how our proposed system helps developers fix vulnerabilities in source code. They were given the same five functions the controlled group was given in Subsection \ref{moti_2}. Our goal is to compare the performance between the controlled and assisted in repairing the vulnerability given the outcome from SOTA methods and the outcome from our system. We provided the participants with the outcome from our model based on the given vulnerable code. Our model's outcome includes the vulnerability classification with a short description, the vulnerable lines, and the root cause of the vulnerability. Figure \ref{fig:survey_our} shows a sample we provided to the participants. We asked the participants to repair the code they were given with the assistance of the outcome provided by our model.

From our analysis in Table \ref{tab:qual_com_1}, we found that 13\% of the participants use ChatGPT extensively for code repair. While we assumed this would be a plausible event, we still considered the outcome generated by ChatGPT in our results. By manually analyzing the written code from these 13\% of the participants, we found that 18\% of the code generated by ChatGPT has vulnerabilities. From this table, we can see that for each CWE, the percentage of assisted participants repairing the code compared to the control group is significantly higher for all cases. The "Repair Improvement" column from Table \ref{tab:qual_com_1} shows the improvement percentage of the participants who could repair the vulnerable code compared to the control group.

Furthermore, we also did a time analysis as a part of our survey. In the given code, for each function the participants have written, we asked them to mention the starting and ending time it took them to write the code. Table \ref{tab:qual_com_1} also shows that, there is a good improvement in the percentage of the assisted participants who solved the vulnerability more quickly. For example, for CWE-20, CWE-264, and CWE-125, the participants took at least 10\% less time than their controlled counterparts. This improvement in time shows that our model can significantly boost a developer's time to repair a vulnerability by providing a root cause of the vulnerability and a static description of the CWE.

\begin{figure}[!ht]
    %\begin{framed}
        \centering
        
        \includegraphics[width=0.5\textwidth]{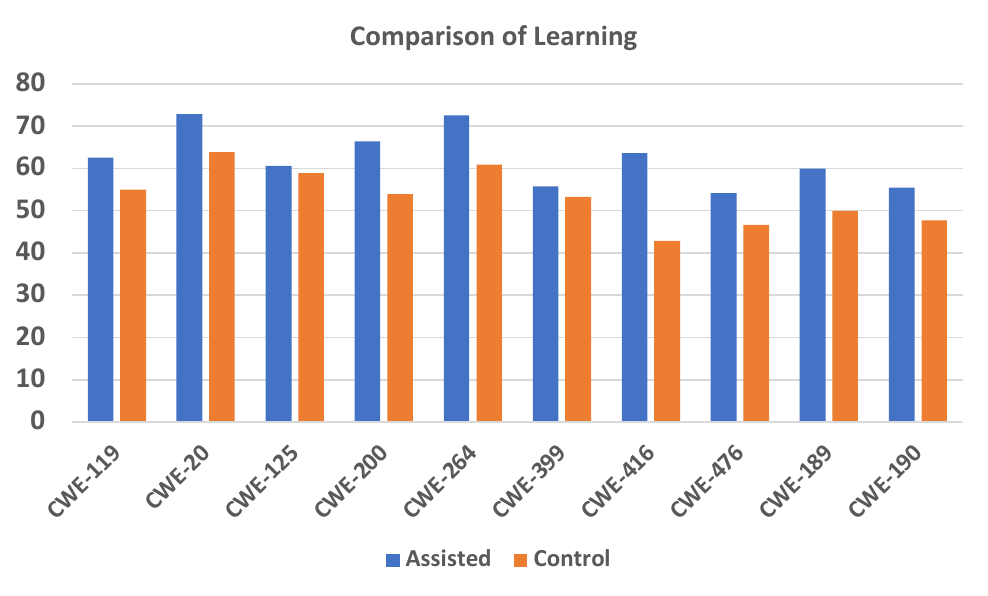}
        
    %\end{framed}
    
    \caption{Performance of Developer Education when Comparing Assisted vs. Control Group}
    \label{fig:learning_1}
\end{figure}

\paragraph{\textbf{Developer Education}} In this part, we use the control and assisted group participants to determine how our system eventually educates developers to write vulnerability-free code. Furthermore, the control group has never seen an outcome from our model. Therefore, we hypothesized that by seeing the outcome from our model, the assisted groups have developed the ability to write more vulnerability-free code. In other words, we want to determine how self-sufficient the developers have become in fixing code vulnerabilities entirely.

To test this, we asked both groups of developers with the same ten functions we asked them to reuse the code they completed in Subsection \ref{moti_1}. Then, we asked both groups to check their code, and based on their learning from the previous phase, they could fix any existing vulnerabilities in their code without any assistance.

Our comparison from Figure \ref{fig:learning_1} shows that the participants from the assisted group have a higher success rate in finding and repairing vulnerabilities from their original code across all CWE types. Furthermore, in the controlled group, there were still vulnerabilities in 49\% of the codes they submitted after resubmission. However, in the assisted group, we see only 38\% of the code has a vulnerability. Hence, we see an almost 9\% improvement in developer learning from assisted groups when using our system compared to SOTA methods. Furthermore, after the survey, 42.3\% of the assisted participants responded that they feel more comfortable with analyzing and repairing source code vulnerability, compared to 34\% from the control group.
\color{black}

\begin{table}[b]
\centering

\caption{Vulnerability classification and localization with Big-Vul and D2A dataset}
%\vspace{0.3cm}
\begin{tabular}{p{0.025\textwidth}
                 p{0.08\textwidth}
                 p{0.030\textwidth}
                 p{0.006\textwidth}
                 p{0.006\textwidth}
                 p{0.006\textwidth}
                 p{0.006\textwidth}}

\toprule
\textbf{Data}                    & \textbf{Model}          & \multicolumn{1}{l}{\textbf{IoU}} &  \multicolumn{1}{l}{\textbf{Acc.}} & \multicolumn{1}{l}{\textbf{F1}} & \multicolumn{1}{l}{\textbf{Pre.}} & \multicolumn{1}{l}{\textbf{Rec.}} \\ \midrule
\multirow{4}{*}{\rotatebox{90}{D2A}}    & Devign         & 0.58                                                 & -                        & -                      & -                        & -                        \\
                        & VELVET         & 0.55                                               & 0.59                     & 0.58                   & \textbf{0.70}            & 0.50                     \\
                        & LineVul        & 0.42                                               & -                        & -                      & -                        & -                        \\
                        & PFGCN & -                & 0.61                         & 0.61            & 0.61          & 0.62                           \\ 
                        
                        & \textbf{Ours} & \textbf{0.72}                                      & \textbf{0.62}            & \textbf{0.68}          & 0.60                     & \textbf{0.65}            \\
                        \midrule

\multirow{5}{*}{\rotatebox{90}{BigVul}} & VulChecker         & 0.44                                                 & -                        & 0.26                   & 0.18                     & 0.52                     \\
                        & VELVET         & 0.45                                                 & -                        & -                      & -                        & -                        \\
                        & LineVul        & 0.45                                                 & -                        & 0.56                   & \textbf{0.66}            & 0.60                     \\
                        & IVDetect       & -                                                       & -                        & 0.35                   & 0.23                     & 0.72                     \\
                        & \textbf{Ours} & \textbf{0.49}                                         & \textbf{0.65}            & \textbf{0.62}          & 0.62                     & \textbf{0.66}            \\ \bottomrule
\end{tabular}
\label{tab:class}
\end{table}

\subsection{RQ2: Quantitative Analysis}
\label{rq2}

%\textbf{Is it possible to identify the source codes' root causes of vulnerability in addition to classifying, describing and localizing them?}

To quantitatively demonstrate the effectiveness of our vulnerability classifier and find the vulnerable line, we used our proposed  T5-GCN model. The task is to classify with an added static description initially and then localize the vulnerable line.

%\vspace{-2mm}
\paragraph{\textbf{Vulnerability Classification}}
Initially, we classified vulnerability using our proposed T5-GCN. For the D2A dataset, we compared our model with Devign \cite{ding2021velvet}, VELVET \cite{ding2021velvet}, and LineVul \cite{fu2022linevul} and PFGCN \cite{islam2023unbiased}. For the Big-Vul dataset, we compared our model with VulChecker \cite{mirskyvulchecker}, VELVET \cite{ding2021velvet}, LineVul \cite{fu2022linevul}, and IVDetect \cite{li2021vulnerability}. For the D2A dataset, our model achieves almost a 3\% increase in accuracy; however, 10\% in F1 score and 15\% in recall. For the BigVul dataset, we used the results shared by \cite{fu2022linevul}. Numerical comparison with previous models for the BigVul dataset shows that our model significantly improves F1 and recall. Table \ref{tab:class} shows a detailed comparison of the classification of our T5-GCN model with other models for the two datasets. Since the D2A dataset has no classification information, accuracy, F1, precision, and recall are based on the detection task.

\begin{figure}[t]
    %\begin{framed}
        \centering
        
        \includegraphics[width=0.5\textwidth]{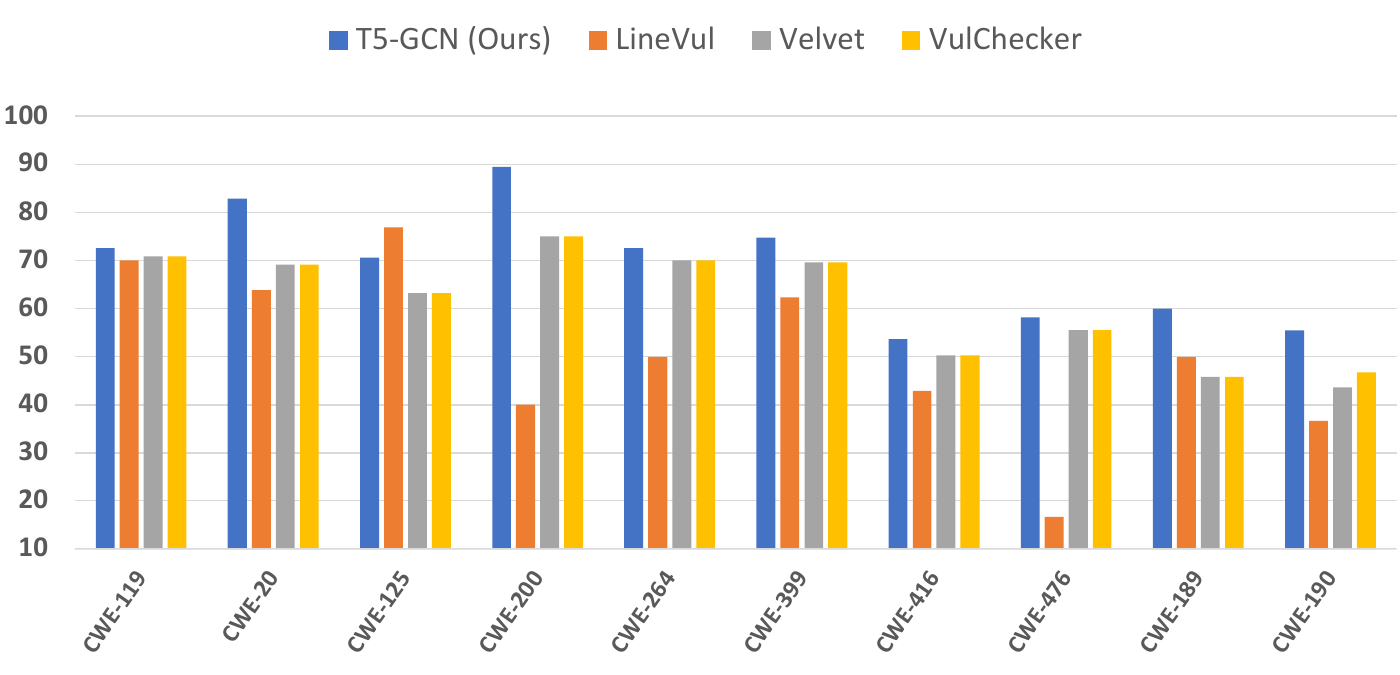}
        
    %\end{framed}
    
    \caption{Multi class Vulnerability Classification in Comparison of SOTA Methods to our proposed T5-GCN. X-axis is the vulnerability category, and Y-axis demonstrates the F1 Score}
    \label{fig:mult_chart}
\end{figure}

To assess the performance of vulnerability classification using our model, we tasked it to classify 10 categories of vulnerabilities as demonstrated in Figure \ref{fig:mult_chart}, which frequently appear in the top 25 vulnerable categories in CWE \cite{cwe}, as mentioned by \cite{fan2020ac}. 

%For multiclass vulnerability detection, our goal was to detect the vulnerability category with a CWE number. If the code is not vulnerable, the model outputs 0. After determining the vulnerable category, the next task is to localize the vulnerability.

Figure \ref{fig:mult_chart} compares the classification F1 score 
for each CWE category. Furthermore, from our training dataset, we found that CWE-119, CWE-20, and CWE-200 comparatively have the highest number of training samples, having more than 1000 examples. Consequently, CWE-416, CWE-476, and CWE-190 have the least training examples, with less than 400 training samples. Therefore, we see that CWE-119 and CWE-20 achieve an F1 score of 70\% and 63.83\%, respectively. The classification capability of our model highly outnumbers other SOTA models like VulChecker \cite{mirskyvulchecker}, VELVET \cite{ding2021velvet}, and LineVul \cite{fu2022linevul} and PFGCN \cite{islam2023unbiased}. However, for other classes like CWE-416, CWE-476, and CWE-190, CWE-190 we see a moderate downfall in performance due to small but still achieving higher performance than SOTA models due to the use of Focal Loss during the training phase of the model.

%In order to analyze the performance of our multitask vulnerability description model, our model was tasked with classifying 40 categories of vulnerability from VulF dataset where each category is associated with a CWE number and a description. For multiclass vulnerability classification, the goal of the classifier initially is to detect whether vulnerability exists in the code, and if a vulnerability exists, provide the CWE number as depicted in Figure \ref{fig:code}. This experiment tests our classifier's ability to detect and classify vulnerable code samples. Table \ref{fig:classification} compares the classification performance of various models against ours when tested with different datasets. Table \ref{tab:multiclassvuln} shows that our VulF dataset is mildly imbalanced. For example, the number of vulnerable codes for CWE-676, CWE-362, and CWE-662 is deficient, less than 300. On the other hand, there are 4713 code samples for CWE-119 and 1380 for CWE-704. We prevented training a biased model by using the Focal Loss function during the model's training.

\paragraph{\textbf{Finding Vulnerable Line}} Based on the prediction by our classifier, we pass the identified vulnerable functions to our vulnerability localizer. Table \ref{tab:class} shows that our localizer produced an IoU of 0.72 and 0.49 IoU, respectively, for D2A and BigVul datasets. Our model produced at least 14\% higher IoUs compared to VulChecker \cite{mirskyvulchecker} on the D2A dataset and at least 4\% higher on the BigVul Dataset. We compared our model with other models like VELVET \cite{ding2021velvet} and LineVul \cite{fu2022linevul}. We observe that our model with SVG as an input graph produced a significantly higher IoU score in all the cases.

\begin{table}[b]
\centering
\caption{Generalizability Testing of our Model. We test with out-of-sample data during evaluation to test the generalizability of the model}
%\vspace{0.3cm}
\label{tab:generaility}
\begin{tabular}{p{0.08\textwidth}
                 p{0.07\textwidth}
                 p{0.03\textwidth}
                 p{0.03\textwidth}
                 p{0.03\textwidth}
                 p{0.03\textwidth}
                 p{0.04\textwidth}}
\toprule
\textbf{Training Data} & \textbf{Evaluation Data} & \textbf{IoU} &\textbf{Acc.} & \textbf{F1} & \textbf{Pre.} & \textbf{Rec.}      \\ \midrule

\multirow{2}{*}{D2A} & D2A & 0.72 & 0.62  & 0.68  &0.68  & 0.65   \\
                                    & BigVul & 0.40 & 0.52  & 0.55  &0.57  & 0.52 \\ \midrule
                                                    
\multirow{2}{*}{BigVul} & BigVul & 0.49 & 0.65  & 0.62  &  0.66  & 0.66     \\
                                    & D2A & 0.44 & 0.57  & 0.59  &0.60  & 0.51 \\ \midrule
                                    
\multirow{3}{*}{Combined} & D2A & 0.74 & 0.66  & 0.75  &0.70  & 0.70     \\

& BigVul & 0.50 & 0.64  & 0.66  &0.67  & 0.70 \\

& Combined & 0.65 & 0.68  & 0.64  &0.66  & 0.69 
                                    
%\vspace{-5mm}                                  
\\\bottomrule

\end{tabular}
\end{table}

\begin{table}[t]
\caption{N-day Vulnerability Discovery and Localization from IoT OS Repositories}
\centering
\begin{tabular}{l|r|r}
\hline
IoT OS                           & N-Day Vulnerability & Count \\ \hline
TinyOS                           & N/A                 & 0     \\ \hline
\multirow{2}{*}{Contiki}         & CWE-119             & 4     \\
                                 & CWE-189              & 1     \\ \hline
\multirow{3}{*}{Zephyr}          & CWE-264             & 2     \\
                                 & CWE-119             & 4     \\
                                 & CWE-399             & 1     \\ \hline
\multirow{2}{*}{FreeRTOS}        & CWE-119             & 1     \\
                                 & CWE-190             & 2     \\ \hline
\multirow{2}{*}{RIOT-OS}         & CWE-119             & 5     \\
                                 & CWE-476             & 1     \\ \hline
\multirow{2}{*}{Raspberry-Pi OS} & CWE-200             & 2     \\
                                 & CWE-119             & 1     \\ \hline
\end{tabular}
\label{tab:iot_data}
\end{table}

%\vspace{0.2cm}

\subsection{RQ3: Generalizability Testing}
\label{rq3}

%\textbf{Does our root cause detection system identify 0 and N-day vulnerabilities from the wild?}

\textbf{Generalizability Testing:} We test our model's generalizability by testing the model's outcome on IoT device operating systems. To achieve this, we trained our model with one dataset and tested the model’s performance on a different dataset to observe our model’s generalization efforts. Furthermore, we combined the D2A and BigVul datasets to achieve more generalizability and tested the model's performance. When we trained the model on the D2A dataset and tested it on the BigVul dataset and vice versa, we can see from Table \ref{tab:generaility} that the accuracy has decreased. However, when we trained the model with the combined dataset, we saw that the F1 scores had increased up to 7\% and the IoU score had increased by 2\% for D2A dataset. Moreover, the IoU score has increased 1\% for BigVul Dataset. We observe that when we use the combined dataset for training, metrics for D2A increase significantly since D2A dataset is 26 times smaller than BigVul, hence providing more generalizability to our model.

\textbf{N-day and Zero-day Analysis:} We implement a vulnerability localization workflow on six  C/C++ based IoT operating system repositories from GitHub to discover wild N-day vulnerabilities. Initially, we scan the repositories using JOERN to extract the functions. Table \ref{tab:iot_data} shows that from the six repositories, we could detect 24 vulnerabilities; however, we could not detect any N-day vulnerabilities from TinyOS. Moreover, from Table \ref{tab:iot_data}, we can see that CWE200 and CWE-120 are spread along multiple repositories, providing us with an insight into what kind of vulnerabilities are prevalent 
across IoT devices.

We found three zero-day vulnerable code samples from the IoT device operating systems. Our model was able to classify and find the vulnerable line of these three zero-day samples. In this case, our model was also trained on the combined dataset consisting of D2A and BigVul. We verified the outcome of our model with security experts. Out of these three confirmed zero-day vulnerabilities, two of them were from Zephyr, and one was from FreeRTOS. All of the vulnerabilities we categorized as CWE-119.

%%%%%%%%%%%%%%%%%%%%%%%%%%%%%%%%%%%%

%\vspace{0.2cm}
%\textbf{RQ3: Can we find zero-day vulnerabilities from program %samples in IoT operating systems?}
%\vspace{0.2cm}

%\vspace{-0.4cm}

%\textbf{Zero-day Analysis:} We found three zero-day vulnerable code samples from the IoT device operating systems. Our model was able to classify and find the vulnerable line of these three zero-day samples. In this case, our model was also trained on the combined dataset consisting of D2A and BigVul. We verified the outcome of our model with security experts. Out of these three confirmed zero-day vulnerabilities, two of them were from Zephyr, and one was from FreeRTOS. All of the vulnerabilities we categorized as CWE-119.

%\vspace{-2mm}
%%%%%%%%%%%%%

%\subsection{Vulnerable IoT Code Collection  (Dylan)}
%We implemented a web cralwer to collect vulnerable code samples from GitHub Advisories, a database on GitHub that provides a centralized location for security advisories related to open source software. We collected 11 samples from trending IoT repositories including TinyOS, Contiki, FreeRTOS, RIOT-OS, Zephyr, and RaspberryPi. Each of the 11 samples has a severity rating of high or critical and the majority of the advisories are related to memory errors that could allow attackers to execute remote code or access sensitive information on the device.

%\textbf{Case Study 1}:

%%%%%%%%%%%%%

% Repository Name| Number of Files | Number of Lines | Number of Functions | Func call per functions | # External Calls

\section{Related Work}
\label{3_related_work}
%This discussion offers a thorough review of key advancements in IoT Source Code Vulnerabilities, Vulnerability Detection, Source Code Vulnerability Localization, and Root Cause Analysis and Explainability. 

A recent study done by Werkme \cite{wermke2022committed} suggests that addressing vulnerabilities constitutes a noteworthy issue in open-source software (OSS) because many maintainers ignore minor security vulnerabilities and only address significant concerns as they are brought up by the community or by automated security software. One such automated security software, V1SCAN, was developed by Woo \cite{woo2023v1scan} to scan for OSS reuse, a common vulnerability introduced in OSS when maintainers copy out-of-date vulnerable OSS code.  

In this section we focus on identifying areas where current research falls short in analyzing vulnerabilities in OSS and setting the stage for our new approach. Current detection tools such as Devign \cite{zhou2019devign}, VELVET \cite{ding2021velvet}, and LineVul \cite{fu2022linevul} are helpful in identifying source code Vulnerabilities. However, they don't offer enough feedback to guide developers in fully addressing these vulnerabilities.

\paragraph{\textbf{Advancements in Vulnerability Detection}}
Vulnerability detection has progressed from conventional machine learning (ML) techniques to more flexible and universally applicable deep learning-based solutions. As Lin et al. \cite{lin2020software} elucidates, ML methods offer a promising avenue for automated vulnerability discovery. Simultaneously, deep learning-based solutions such as VulDeepecker \cite{li2018vuldeepecker} and $\mu$VulDeepecker \cite{zou2019mu} offer universal applicability; nevertheless, they presented a heavy reliance on feature engineering for vulnerability detection. The advancements of Devign \cite{zhou2019devign} contrasted these approaches compared to VulBERTa \cite{hanif2022vulberta}, and ReVEAL \cite{chakraborty2021deep}, which used Code Property Graph (CPG) \cite{yamaguchi2014modeling} based techniques to identify vulnerabilities, offering improved flexibility. Alongside these, Abstract Syntax Tree (AST) based techniques were proposed by Bilgin et al. \cite{bilgin2020vulnerability} and others \cite{lin2019software}, \cite{li2021towards}, \cite{dam2018automatic}, to retain the syntactic information of the source code during detection. Furthermore, transformer-based models like RoBERTa \cite{liu2019roberta} were adopted by VulBERTa \cite{hanif2022vulberta} and \cite{thapa2022transformer} to detect vulnerabilities in source code. Lastly, Islam et al. \cite{islam2023unbiased} proposed a semantic understanding of programming languages to classify and detect vulnerabilities.

\paragraph{\textbf{Methods of Source Code Vulnerability Localization}}
The task of localizing vulnerabilities within source code has seen the application of various methods, from traditional rule-based static analysis tools to innovative deep learning techniques. While tools like Cppcheck \cite{cppcheck}, FlawFinder \cite{flawfinder}, RATS \cite{rats}, and Infer \cite{infer} provide direct approaches to vulnerability localization, their high false positive and false negative rates \cite{yamaguchi2015pattern} underscore the need for more reliable methods. A promising avenue is provided by the use of deep learning, as evidenced by the fine-grained vulnerability detection and locator systems proposed by Vuldeelocator \cite{li2021vuldeelocator} and DeepLineDP \cite{pornprasit2022deeplinedp}, leveraging bidirectional Recurrent Neural Networks (RNN). Additionally, the ensemble graph-transformer learning approach by VELVET \cite{ding2021velvet} and the non-conventional explainability technique by LineVul \cite{fu2022linevul} demonstrate advances in detecting and localizing vulnerabilities at the statement level. VulChecker \cite{mirskyvulchecker} presents a unique approach, using an intermediate representation called LLVM for vulnerability localization.

\paragraph{\textbf{Challenges and Evolution of Root Cause Analysis}}
Identifying the root cause of vulnerabilities within a system is a formidable task, especially given the disconnect between states where code crashes and actual root causes. Over time, several notable approaches have emerged. AutoPaG \cite{lin2007autopag}, for instance, leveraged the data flow of programs to analyze vulnerabilities such as out-of-bound, buffer overflow errors, and general boundary condition errors. To isolate the root cause of thread scheduling issues, Choi et al. \cite{choi2002isolating} proposed narrowing down differences in thread schedules, resulting in the failure of the program. Further, Failure Sketching \cite{kasikci2015failure} introduced a cooperative adaptation of static and dynamic analysis for identifying root causes of production failures. More recent contributions, such as ARCUS \cite{yagemann2021arcus} and Yagemann et al.'s methodology \cite{yagemann2021automated}, automate root cause analysis through execution flags and binary level analysis, and tracking execution traces, respectively, enhancing the detection of issues like buffer overflow and use-after-free exploits.

\paragraph{\textbf{Explainability for Development-Phase Vulnerability Analysis}} While the aforementioned methods focus on vulnerability analysis during a program's execution, it is equally critical, from a developer's perspective, to scrutinize vulnerabilities during the development phase \cite{islam2023unbiased} \cite{braz2022less}. This has led to the proposal of explainable techniques that pinpoint relevant features contributing to a program's vulnerability \cite{mosolygo2021towards}, \cite{sotgiu2022explainability}, \cite{mao2020explainable}. Asm2Seq \cite{taviss2023asm2seq} and VulANalyzeR \cite{li2023vulanalyzer} took this concept further by introducing explainable deep learning-based approaches for identifying binary vulnerabilities in source code. Notably, VulANalyzeR \cite{li2023vulanalyzer} employed an attention-based explainable mechanism to unearth the root cause of vulnerabilities. 

%\mj{One summary paragraph here on how our proposed work is different from works so far, and how it helps to fill a critical research gap, would be useful for the reader to understand the novely of this paper!}

To conclude, we have explored various methods for vulnerability analysis, highlighting the shift towards development-phase vulnerability analysis and the use of explainable techniques. However, none of the current methods tries to identify the root cause of the vulnerability while ensuring the usability of their proposed system. Therefore, our research aims to fill the existing gap in finding the root cause of the vulnerability of code with a proper, informative, and usable system.

%To summarize, our exploration of different methods of vulnerability analysis shows a shift towards examining vulnerabilities during development, as well as the increased use of explainable techniques. However, there is still a gap in token-based vulnerability localization, especially for the C/C++ programming languages. This gap is what our research aims to fill. With the aid of Large Language Models, we provide situational examples to better understand the source of vulnerabilities, leading to a shift from simply identifying vulnerabilities to creating effective strategies to rectify them. The details of our proposed approach, designed to significantly improve the handling of Source Code Vulnerabilities, are discussed further in the following sections.

%\color{red}
%\section{Limitations}
%\color{black}

\section{Conclusion}
\label{8_conclusion}

This study presents a comprehensive end-to-end solution to address the lack of a usable vulnerability analysis system by finding its root cause using an explainable technique. Initially, we conduct survey analyses to demonstrate the challenges a developer faces from the SOTA methods. We further interviewed the developers to identify the gaps that must be addressed to mitigate the current challenges.
We also present classification, localization, and a short description to aid the developer in repairing vulnerabilities. Our approach leverages the combined power of LLM CodeT5 and a GCN in a multitask setting to analyze source code. Our proposed model accurately classifies vulnerability categories, localizes vulnerable lines, and identifies underlying causes across multiple datasets. Furthermore, we test our system's capability by comparing our outcome's effectiveness vs. the outcome from the SOTA methods. Our model exhibits generalizability by detecting N-day and zero-day vulnerabilities in IoT operating system source code. 
Although, in our survey, we tried to the best of our ability to provide a realistic scenario for code development, this needs to reflect the development condition in a real-life situation holistically. Development in real-life conditions constitutes a complex environment with multiple developers writing code in multiple files, which further increases the chances of vulnerability. In our future work, we aim to extend our methodology by extending the root cause capabilities of our model and combining it with generative models to suggest a possible fix for the developers.

\bibliographystyle{unsrt}
\bibliography{main}

\begin{thebibliography}{10}

\bibitem{al2022idetect}
Abdullah Al-Boghdady, Mohammad El-Ramly, and Khaled Wassif.
\newblock idetect for vulnerability detection in internet of things operating systems using machine learning.
\newblock {\em Scientific Reports}, 12(1):1--12, 2022.

\bibitem{ladisa2023sok}
Piergiorgio Ladisa, Henrik Plate, Matias Martinez, and Olivier Barais.
\newblock Sok: Taxonomy of attacks on open-source software supply chains.
\newblock In {\em 2023 IEEE Symposium on Security and Privacy (SP)}, pages 1509--1526. IEEE, 2023.

\bibitem{peisert2021perspectives}
Sean Peisert, Bruce Schneier, Hamed Okhravi, Fabio Massacci, Terry Benzel, Carl Landwehr, Mohammad Mannan, Jelena Mirkovic, Atul Prakash, and James~Bret Michael.
\newblock Perspectives on the solarwinds incident.
\newblock {\em IEEE Security \& Privacy}, 19(2):7--13, 2021.

\bibitem{log4j}
Log4j, https://nvd.nist.gov/vuln/detail/CVE-2021-44228.

\bibitem{synopsysrisc}
Synopsys.
\newblock Open source security and risk analysis report.
\newblock 2023.

\bibitem{fedreg}
US~Government.
\newblock Federal register.
\newblock 2023.

\bibitem{khoury2023secure}
Rapha{\"e}l Khoury, Anderson~R Avila, Jacob Brunelle, and Baba~Mamadou Camara.
\newblock How secure is code generated by chatgpt?
\newblock {\em arXiv preprint arXiv:2304.09655}, 2023.

\bibitem{sandoval2022security}
Gustavo Sandoval, Hammond Pearce, Teo Nys, Ramesh Karri, Brendan Dolan-Gavitt, and Siddharth Garg.
\newblock Security implications of large language model code assistants: A user study.
\newblock {\em arXiv preprint arXiv:2208.09727}, 2022.

\bibitem{democr1}
Jukka Niiranen.
\newblock Democratizing code, https://jukkaniiranen.com/2021/04/democratizing-code/.

\bibitem{democr2}
AKILEK~Akilek Consulting.
\newblock Democratizing programming: How ai enables everyone to become a programmer, https://www.linkedin.com/pulse/democratizing-programming-how-ai-enables-everyone-become/.

\bibitem{pearce2022asleep}
Hammond Pearce, Baleegh Ahmad, Benjamin Tan, Brendan Dolan-Gavitt, and Ramesh Karri.
\newblock Asleep at the keyboard? assessing the security of github copilot’s code contributions.
\newblock In {\em 2022 IEEE Symposium on Security and Privacy (SP)}, pages 754--768. IEEE, 2022.

\bibitem{green2016developers}
Matthew Green and Matthew Smith.
\newblock Developers are not the enemy!: The need for usable security apis.
\newblock {\em IEEE Security \& Privacy}, 14(5):40--46, 2016.

\bibitem{acar2016you}
Yasemin Acar, Sascha Fahl, and Michelle~L Mazurek.
\newblock You are not your developer, either: A research agenda for usable security and privacy research beyond end users.
\newblock {\em 2016 IEEE Cybersecurity Development (SecDev)}, pages 3--8, 2016.

\bibitem{assal2019think}
Hala Assal and Sonia Chiasson.
\newblock 'think secure from the beginning' a survey with software developers.
\newblock In {\em Proceedings of the 2019 CHI conference on human factors in computing systems}, pages 1--13, 2019.

\bibitem{weir2020needs}
Charles Weir, Ben Hermann, and Sascha Fahl.
\newblock From needs to actions to secure apps? the effect of requirements and developer practices on app security.
\newblock In {\em 29th USENIX security symposium (USENIX security 20)}, pages 289--305, 2020.

\bibitem{zhou2019devign}
Yaqin Zhou, Shangqing Liu, Jingkai Siow, Xiaoning Du, and Yang Liu.
\newblock Devign: Effective vulnerability identification by learning comprehensive program semantics via graph neural networks.
\newblock {\em Advances in neural information processing systems}, 32, 2019.

\bibitem{li2018vuldeepecker}
Zhen Li, Deqing Zou, Shouhuai Xu, Xinyu Ou, Hai Jin, Sujuan Wang, Zhijun Deng, and Yuyi Zhong.
\newblock Vuldeepecker: {A} deep learning-based system for vulnerability detection.
\newblock In {\em 25th Annual Network and Distributed System Security Symposium, {NDSS} 2018, San Diego, California, USA, February 18-21, 2018}. The Internet Society, 2018.

\bibitem{li2021vuldeelocator}
Zhen Li, Deqing Zou, Shouhuai Xu, Zhaoxuan Chen, Yawei Zhu, and Hai Jin.
\newblock Vuldeelocator: a deep learning-based fine-grained vulnerability detector.
\newblock {\em IEEE Transactions on Dependable and Secure Computing}, 2021.

\bibitem{nguyen2021regvd}
Van-Anh Nguyen, Dai~Quoc Nguyen, Van Nguyen, Trung Le, Quan~Hung Tran, and Dinh Phung.
\newblock Re{GVD}: Revisiting graph neural networks for vulnerability detection.
\newblock In {\em Deep Learning for Code Workshop}, 2022.

\bibitem{islam2023unbiased}
N.~Islam, G.~Parra, D.~Manuel, E.~Bou-Harb, and P.~Najafirad.
\newblock An unbiased transformer source code learning with semantic vulnerability graph.
\newblock In {\em 2023 IEEE 8th European Symposium on Security and Privacy (EuroS\&P)}, pages 144--159, Los Alamitos, CA, USA, jul 2023. IEEE Computer Society.

\bibitem{guo2020graphcodebert}
Daya Guo, Shuo Ren, Shuai Lu, Zhangyin Feng, Duyu Tang, Shujie Liu, Long Zhou, Nan Duan, Alexey Svyatkovskiy, Shengyu Fu, Michele Tufano, Shao~Kun Deng, Colin~B. Clement, Dawn Drain, Neel Sundaresan, Jian Yin, Daxin Jiang, and Ming Zhou.
\newblock Graphcodebert: Pre-training code representations with data flow.
\newblock {\em CoRR}, abs/2009.08366, 2020.

\bibitem{pearce2023examining}
Hammond Pearce, Benjamin Tan, Baleegh Ahmad, Ramesh Karri, and Brendan Dolan-Gavitt.
\newblock Examining zero-shot vulnerability repair with large language models.
\newblock In {\em 2023 IEEE Symposium on Security and Privacy (SP)}, pages 2339--2356. IEEE, 2023.

\bibitem{joshi2023repair}
Harshit Joshi, Jos{\'e}~Cambronero Sanchez, Sumit Gulwani, Vu~Le, Gust Verbruggen, and Ivan Radi{\v{c}}ek.
\newblock Repair is nearly generation: Multilingual program repair with llms.
\newblock In {\em Proceedings of the AAAI Conference on Artificial Intelligence}, volume~37, pages 5131--5140, 2023.

\bibitem{chen2022neural}
Zimin Chen, Steve Kommrusch, and Martin Monperrus.
\newblock Neural transfer learning for repairing security vulnerabilities in c code.
\newblock {\em IEEE Transactions on Software Engineering}, 49(1):147--165, 2022.

\bibitem{2020t5}
Colin Raffel, Noam Shazeer, Adam Roberts, Katherine Lee, Sharan Narang, Michael Matena, Yanqi Zhou, Wei Li, and Peter~J. Liu.
\newblock Exploring the limits of transfer learning with a unified text-to-text transformer.
\newblock {\em Journal of Machine Learning Research}, 21(140):1--67, 2020.

\bibitem{wang2021codet5}
Yue Wang, Weishi Wang, Shafiq Joty, and Steven~CH Hoi.
\newblock Codet5: Identifier-aware unified pre-trained encoder-decoder models for code understanding and generation.
\newblock {\em arXiv preprint arXiv:2109.00859}, 2021.

\bibitem{touvron2023llama}
Hugo Touvron, Thibaut Lavril, Gautier Izacard, Xavier Martinet, Marie-Anne Lachaux, Timoth{\'e}e Lacroix, Baptiste Rozi{\`e}re, Naman Goyal, Eric Hambro, Faisal Azhar, et~al.
\newblock Llama: Open and efficient foundation language models.
\newblock {\em arXiv preprint arXiv:2302.13971}, 2023.

\bibitem{zelikman2023self}
Eric Zelikman, Eliana Lorch, Lester Mackey, and Adam~Tauman Kalai.
\newblock Self-taught optimizer (stop): Recursively self-improving code generation.
\newblock {\em arXiv preprint arXiv:2310.02304}, 2023.

\bibitem{bellare1997forward}
Mihir Bellare and Bennet Yee.
\newblock Forward integrity for secure audit logs.
\newblock Technical report, Citeseer, 1997.

\bibitem{capobianco2019employing}
Frank Capobianco, Rahul George, Kaiming Huang, Trent Jaeger, Srikanth Krishnamurthy, Zhiyun Qian, Mathias Payer, and Paul Yu.
\newblock Employing attack graphs for intrusion detection.
\newblock In {\em Proceedings of the New Security Paradigms Workshop}, pages 16--30, 2019.

\bibitem{han2020unicorn}
Xueyuan Han, Thomas Pasquier, Adam Bates, James Mickens, and Margo Seltzer.
\newblock Unicorn: Runtime provenance-based detector for advanced persistent threats.
\newblock {\em arXiv preprint arXiv:2001.01525}, 2020.

\bibitem{ji2017rain}
Yang Ji, Sangho Lee, Evan Downing, Weiren Wang, Mattia Fazzini, Taesoo Kim, Alessandro Orso, and Wenke Lee.
\newblock Rain: Refinable attack investigation with on-demand inter-process information flow tracking.
\newblock In {\em Proceedings of the 2017 ACM SIGSAC conference on computer and communications security}, pages 377--390, 2017.

\bibitem{kang2011dta}
Min~Gyung Kang, Stephen McCamant, Pongsin Poosankam, and Dawn Song.
\newblock Dta++: dynamic taint analysis with targeted control-flow propagation.
\newblock In {\em NDSS}, 2011.

\bibitem{alenezi2020relationship}
Mamdouh Alenezi and Mohammad Zarour.
\newblock On the relationship between software complexity and security.
\newblock {\em arXiv preprint arXiv:2002.07135}, 2020.

\bibitem{infer}
Infer.
\newblock Infer.
\newblock 2013.

\bibitem{cppcheck}
Cppcheck.
\newblock https://cppcheck.sourceforge.io/.
\newblock 2022.

\bibitem{fu2022linevul}
Michael Fu and Chakkrit Tantithamthavorn.
\newblock Linevul: A transformer-based line-level vulnerability prediction.
\newblock 03 2022.

\bibitem{pornprasit2022deeplinedp}
Chanathip Pornprasit and Chakkrit Tantithamthavorn.
\newblock Deeplinedp: Towards a deep learning approach for line-level defect prediction.
\newblock {\em IEEE Transactions on Software Engineering}, 2022.

\bibitem{mirskyvulchecker}
Yisroel Mirsky, George Macon, Michael Brown, Carter Yagemann, Matthew Pruett, Evan Downing, Sukarno Mertoguno, and Wenke Lee.
\newblock Vulchecker: Graph-based vulnerability localization in source code.

\bibitem{fu2022vulrepair}
Michael Fu, Chakkrit Tantithamthavorn, Trung Le, Van Nguyen, and Dinh Phung.
\newblock Vulrepair: a t5-based automated software vulnerability repair.
\newblock In {\em Proceedings of the 30th ACM Joint European Software Engineering Conference and Symposium on the Foundations of Software Engineering}, pages 935--947, 2022.

\bibitem{ding2021velvet}
Yangruibo Ding, Sahil Suneja, Yunhui Zheng, Jim Laredo, Alessandro Morari, Gail Kaiser, and Baishakhi Ray.
\newblock Velvet: a novel ensemble learning approach to automatically locate vulnerable statements.
\newblock {\em arXiv preprint arXiv:2112.10893}, 2021.

\bibitem{cwe}
CWE.
\newblock Common weakness enumeration.
\newblock 2022.

\bibitem{sennrich-etal-2016-neural}
Rico Sennrich, Barry Haddow, and Alexandra Birch.
\newblock Neural machine translation of rare words with subword units.
\newblock In {\em Proceedings of the 54th Annual Meeting of the Association for Computational Linguistics (Volume 1: Long Papers)}, pages 1715--1725, Berlin, Germany, August 2016. Association for Computational Linguistics.

\bibitem{huang2019text}
Lianzhe Huang, Dehong Ma, Sujian Li, Xiaodong Zhang, and Houfeng Wang.
\newblock Text level graph neural network for text classification.
\newblock {\em arXiv preprint arXiv:1910.02356}, 2019.

\bibitem{zhang2020every}
Yufeng Zhang, Xueli Yu, Zeyu Cui, Shu Wu, Zhongzhen Wen, and Liang Wang.
\newblock Every document owns its structure: Inductive text classification via graph neural networks.
\newblock In {\em Proceedings of the 58th Annual Meeting of the Association for Computational Linguistics}, pages 334--339, 2020.

\bibitem{lin2017focal}
Tsung-Yi Lin, Priya Goyal, Ross Girshick, Kaiming He, and Piotr Doll{\'a}r.
\newblock Focal loss for dense object detection.
\newblock In {\em Proceedings of the IEEE international conference on computer vision}, pages 2980--2988, 2017.

\bibitem{lundberg2017unified}
Scott~M Lundberg and Su-In Lee.
\newblock A unified approach to interpreting model predictions.
\newblock {\em Advances in neural information processing systems}, 30, 2017.

\bibitem{shrikumar2017learning}
Avanti Shrikumar, Peyton Greenside, and Anshul Kundaje.
\newblock Learning important features through propagating activation differences.
\newblock In {\em International conference on machine learning}, pages 3145--3153. PMLR, 2017.

\bibitem{fan2020ac}
Jiahao Fan, Yi~Li, Shaohua Wang, and Tien~N Nguyen.
\newblock Ac/c++ code vulnerability dataset with code changes and cve summaries.
\newblock In {\em Proceedings of the 17th International Conference on Mining Software Repositories}, pages 508--512, 2020.

\bibitem{zheng2021d2a}
Yunhui Zheng, Saurabh Pujar, Burn Lewis, Luca Buratti, Edward Epstein, Bo~Yang, Jim Laredo, Alessandro Morari, and Zhong Su.
\newblock D2a: a dataset built for ai-based vulnerability detection methods using differential analysis.
\newblock In {\em 2021 IEEE/ACM 43rd International Conference on Software Engineering: Software Engineering in Practice (ICSE-SEIP)}, pages 111--120. IEEE, 2021.

\bibitem{joern}
Joern: The bug hunters workbench.
\newblock https://joern.io/.

\bibitem{rezatofighi2019generalized}
Hamid Rezatofighi, Nathan Tsoi, JunYoung Gwak, Amir Sadeghian, Ian Reid, and Silvio Savarese.
\newblock Generalized intersection over union: A metric and a loss for bounding box regression.
\newblock In {\em Proceedings of the IEEE/CVF conference on computer vision and pattern recognition}, pages 658--666, 2019.

\bibitem{li2021vulnerability}
Yi~Li, Shaohua Wang, and Tien~N Nguyen.
\newblock Vulnerability detection with fine-grained interpretations.
\newblock In {\em Proceedings of the 29th ACM Joint Meeting on European Software Engineering Conference and Symposium on the Foundations of Software Engineering}, pages 292--303, 2021.

\bibitem{wermke2022committed}
Dominik Wermke, Noah W{\"o}hler, Jan~H Klemmer, Marcel Fourn{\'e}, Yasemin Acar, and Sascha Fahl.
\newblock Committed to trust: A qualitative study on security \& trust in open source software projects.
\newblock In {\em 2022 IEEE Symposium on Security and Privacy (SP)}, pages 1880--1896. IEEE, 2022.

\bibitem{woo2023v1scan}
Seunghoon Woo, Eunjin Choi, Heejo Lee, and Hakjoo Oh.
\newblock $\{$V1SCAN$\}$: Discovering 1-day vulnerabilities in reused $\{$C/C++$\}$ open-source software components using code classification techniques.
\newblock In {\em 32nd USENIX Security Symposium (USENIX Security 23)}, pages 6541--6556, 2023.

\bibitem{lin2020software}
Guanjun Lin, Sheng Wen, Qing-Long Han, Jun Zhang, and Yang Xiang.
\newblock Software vulnerability detection using deep neural networks: a survey.
\newblock {\em Proceedings of the IEEE}, 108(10):1825--1848, 2020.

\bibitem{zou2019mu}
Deqing Zou, Sujuan Wang, Shouhuai Xu, Zhen Li, and Hai Jin.
\newblock $\mu$ vuldeepecker: A deep learning-based system for multiclass vulnerability detection.
\newblock {\em IEEE Transactions on Dependable and Secure Computing}, 18(5):2224--2236, 2019.

\bibitem{hanif2022vulberta}
Hazim Hanif and Sergio Maffeis.
\newblock Vulberta: Simplified source code pre-training for vulnerability detection.
\newblock {\em CoRR}, abs/2205.12424, 2022.

\bibitem{chakraborty2021deep}
Saikat Chakraborty, Rahul Krishna, Yangruibo Ding, and Baishakhi Ray.
\newblock Deep learning based vulnerability detection: Are we there yet.
\newblock {\em IEEE Transactions on Software Engineering}, 2021.

\bibitem{yamaguchi2014modeling}
Fabian Yamaguchi, Nico Golde, Daniel Arp, and Konrad Rieck.
\newblock Modeling and discovering vulnerabilities with code property graphs.
\newblock In {\em 2014 IEEE Symposium on Security and Privacy}, pages 590--604. IEEE, 2014.

\bibitem{bilgin2020vulnerability}
Zeki Bilgin, Mehmet~Akif Ersoy, Elif~Ustundag Soykan, Emrah Tomur, Pinar {\c{C}}omak, and Leyli Kara{\c{c}}ay.
\newblock Vulnerability prediction from source code using machine learning.
\newblock {\em IEEE Access}, 8:150672--150684, 2020.

\bibitem{lin2019software}
Guanjun Lin, Jun Zhang, Wei Luo, Lei Pan, Olivier De~Vel, Paul Montague, and Yang Xiang.
\newblock Software vulnerability discovery via learning multi-domain knowledge bases.
\newblock {\em IEEE Transactions on Dependable and Secure Computing}, 18(5):2469--2485, 2019.

\bibitem{li2021towards}
Zhen Li, Jing Tang, Deqing Zou, Qian Chen, Shouhuai Xu, Chao Zhang, Yichen Li, and Hai Jin.
\newblock Towards making deep learning-based vulnerability detectors robust.
\newblock {\em arXiv preprint arXiv:2108.00669}, 2021.

\bibitem{dam2018automatic}
Hoa~Khanh Dam, Truyen Tran, Trang Pham, Shien~Wee Ng, John Grundy, and Aditya Ghose.
\newblock Automatic feature learning for predicting vulnerable software components.
\newblock {\em IEEE Transactions on Software Engineering}, 47(1):67--85, 2018.

\bibitem{liu2019roberta}
Yinhan Liu, Myle Ott, Naman Goyal, Jingfei Du, Mandar Joshi, Danqi Chen, Omer Levy, Mike Lewis, Luke Zettlemoyer, and Veselin Stoyanov.
\newblock Ro{\{}bert{\}}a: A robustly optimized {\{}bert{\}} pretraining approach, 2020.

\bibitem{thapa2022transformer}
Chandra Thapa, Seung~Ick Jang, Muhammad~Ejaz Ahmed, Seyit Camtepe, Josef Pieprzyk, and Surya Nepal.
\newblock Transformer-based language models for software vulnerability detection: Performance, model's security and platforms.
\newblock {\em arXiv preprint arXiv:2204.03214}, 2022.

\bibitem{flawfinder}
Flawfinder.
\newblock https://dwheeler.com/flawfinder/.
\newblock 2002.

\bibitem{rats}
RATS.
\newblock Rats.
\newblock 2023.

\bibitem{yamaguchi2015pattern}
Fabian Yamaguchi.
\newblock Pattern-based vulnerability discovery.
\newblock 2015.

\bibitem{lin2007autopag}
Zhiqiang Lin, Xuxian Jiang, Dongyan Xu, Bing Mao, and Li~Xie.
\newblock Autopag: towards automated software patch generation with source code root cause identification and repair.
\newblock In {\em Proceedings of the 2nd ACM symposium on Information, computer and communications security}, pages 329--340, 2007.

\bibitem{choi2002isolating}
Jong-Deok Choi and Andreas Zeller.
\newblock Isolating failure-inducing thread schedules.
\newblock In {\em Proceedings of the 2002 ACM SIGSOFT international symposium on Software testing and analysis}, pages 210--220, 2002.

\bibitem{kasikci2015failure}
Baris Kasikci, Benjamin Schubert, Cristiano Pereira, Gilles Pokam, and George Candea.
\newblock Failure sketching: A technique for automated root cause diagnosis of in-production failures.
\newblock In {\em Proceedings of the 25th Symposium on Operating Systems Principles}, pages 344--360, 2015.

\bibitem{yagemann2021arcus}
Carter Yagemann, Matthew Pruett, Simon~P Chung, Kennon Bittick, Brendan Saltaformaggio, and Wenke Lee.
\newblock Arcus: Symbolic root cause analysis of exploits in production systems.
\newblock In {\em USENIX Security Symposium}, pages 1989--2006, 2021.

\bibitem{yagemann2021automated}
Carter Yagemann, Simon~P Chung, Brendan Saltaformaggio, and Wenke Lee.
\newblock Automated bug hunting with data-driven symbolic root cause analysis.
\newblock In {\em Proceedings of the 2021 ACM SIGSAC Conference on Computer and Communications Security}, pages 320--336, 2021.

\bibitem{braz2022less}
Larissa Braz, Christian Aeberhard, G{\"u}l {\c{C}}alikli, and Alberto Bacchelli.
\newblock Less is more: supporting developers in vulnerability detection during code review.
\newblock In {\em Proceedings of the 44th International Conference on Software Engineering}, pages 1317--1329, 2022.

\bibitem{mosolygo2021towards}
Bal{\'a}zs Mosolyg{\'o}, Norbert V{\'a}ndor, G{\'a}bor Antal, P{\'e}ter Heged{\H{u}}s, and Rudolf Ferenc.
\newblock Towards a prototype based explainable javascript vulnerability prediction model.
\newblock In {\em 2021 International Conference on Code Quality (ICCQ)}, pages 15--25. IEEE, 2021.

\bibitem{sotgiu2022explainability}
Angelo Sotgiu, Maura Pintor, and Battista Biggio.
\newblock Explainability-based debugging of machine learning for vulnerability discovery.
\newblock In {\em Proceedings of the 17th International Conference on Availability, Reliability and Security}, pages 1--8, 2022.

\bibitem{mao2020explainable}
Yi~Mao, Yun Li, Jiatai Sun, and Yixin Chen.
\newblock Explainable software vulnerability detection based on attention-based bidirectional recurrent neural networks.
\newblock In {\em 2020 IEEE International Conference on Big Data (Big Data)}, pages 4651--4656. IEEE, 2020.

\bibitem{taviss2023asm2seq}
Scarlett Taviss, Steven~HH Ding, Mohammad Zulkernine, Philippe Charland, and Sudipta Acharya.
\newblock Asm2seq: Explainable assembly code functional summary generation for reverse engineering and vulnerability analysis.
\newblock {\em Digital Threats: Research and Practice}, 2023.

\bibitem{li2023vulanalyzer}
Litao Li, Steven~HH Ding, Yuan Tian, Benjamin~CM Fung, Philippe Charland, Weihan Ou, Leo Song, and Congwei Chen.
\newblock Vulanalyzer: Explainable binary vulnerability detection with multi-task learning and attentional graph convolution.
\newblock {\em ACM Transactions on Privacy and Security}, 26(3):1--25, 2023.

\end{thebibliography}

\begin{appendices}
    \section{Survey}

\begin{lstlisting}[style=CStyle, label={lst:lst2}, caption=The included headers\, defined constants\, and defined structures used in the survey.]
#include <stdlib.h>
#include <stdio.h>
#include <string.h>

typedef int bool;

#define TRUE 0
#define FALSE 1

/**
 * User of the grocery store system.
 */
typedef struct
{
    char *username; // Username of the user.
    char *password; // Password of the user.
} User;
\end{lstlisting}

To assess the efficacy of our proposed system, we conducted a comprehensive evaluation involving 56 participants divided into control and assisted groups encompassing diverse educational and professional backgrounds with programming capabilities. The participants engaged in an online survey, wherein the survey's objective was presented as an evaluation of code writing quality. Initially, participants were questioned about their software development proficiency, including the extent of their coding experience and primary programming language expertise.

%%%% Survey work flow commented for testing purpose

\subsection{Survey Workflow}

We initially directed participants to complete ten functions within an incomplete C source code file. These functions were intentionally crafted in a manner where potential vulnerabilities could be introduced if not approached with caution. Eligibility criteria for participation necessitated a fundamental understanding of the C programming language, ensuring that the tasks were within the participants' grasp. It is important to emphasize that the ten functions exclusively revolved around elementary C programming concepts and excluded intricate topics like data structures. In cases where participants did not fulfill any of the functions, they were gracefully guided to the survey's conclusion. This approach aimed to maintain a consistent and relevant experience for all participants. Listing \ref{lst:lst2} shows the header file and the structure we provided to the participants.

\begin{lstlisting}[style=CStyle, label={lst:lst3}, caption=A function to write the updated price of each items to a file]
void exportPrices(char **itemStrings, int numItems, char *filename)
{
    /**
     * Problem:
     *  Export the contents of the grocery store system to a file. Each line of the file should
     * contain one grocery store item.
     *
     * Example:
     *  exportPrices({"Apple - $1.25", "Orange - $0.75"}, 2, "out.txt") the contents of out.txt would
     * look like:
     * Apple - $1.25
     * Orange - $0.75
     *
     * Parameters:
     *  itemStrings: an array of grocery store items formatted as "Item - $X.XX".
     *  numItems: the length of itemStrings.
     *  filename: the name of the output file to export prices to.
     */
    //  --> PARTICIPANT: ENTER HERE TIME STARTED <--
}
//  --> PARTICIPANT: ENTER HERE TIME COMPLETED <--
\end{lstlisting}

Furthermore, \ref{lst:lst3} and \ref{lst:lst4} exemplify some other code samples provided to the participants. These are empty functions with a proper description of the problem, a sample test case of the solution, and a clarification on the return type. Inside the function body, there is an option to put the starting and ending times to measure how long it took them to complete the function.

If participants made partial progress on the coding assignment, a series of inquiries ensued regarding their familiarity with code vulnerabilities and the extent of their formal cybersecurity training. Following the collection of security-related background information, participants were assigned to either "Form A" or "Form B," with "Form A" representing the assisted group and "Form B" serving as the control group. The survey structure was thoughtfully arranged such that every alternate participant received "Form B," a measure intended to achieve an equitable distribution of approximately 50\% between the control and assisted groups.

Irrespective of the assigned form, participants were presented with five distinct C functions, which they were asked to repair. The control group only had visibility to the CWE class of the vulnerability and the vulnerable line, as depicted in Figure \ref{fig:survey_sota}. However, the assisted group was provided with the CWE category of the vulnerability with a static description, the vulnerable line, and the root cause of the vulnerability as depicted in Figure \ref{fig:survey_our}.

Upon completing the vulnerability assessment task, participants were further probed with questions assessing their confidence levels in mitigating code vulnerabilities and whether they sought assistance from ChatGPT for code composition. The survey concluded with the collection of pertinent demographic data from each participant.

\begin{lstlisting}[style=CStyle, label={lst:lst4}, caption=A function that primarily checks if the participant scanned values from a string correctly.]
double extractPrice(char *itemString)
{
    /**
     * Problem:
     *  In the grocery store, prices are stored in the system in the format of "Item - $X.XX".
     * This function should extract the price from the string and return it as a double.
     *
     * Example:
     *  extractPrice("Apple - $1.25") returns 1.25
     *
     * Parameters:
     *  itemString: a grocery store item formatted as "Item - $X.XX".
     *
     * Returns:
     *  the price of the item as a double.
     */
    //  --> PARTICIPANT: ENTER HERE TIME STARTED <--
    return 0.0;
}
//  --> PARTICIPANT: ENTER HERE TIME COMPLETED <--
\end{lstlisting}

    \label{9_appendix}
\end{appendices}

\end{document}